\documentclass[aps,prd,twocolumn,superscriptaddress,preprintnumbers,nofootinbib]{revtex4-1}

\usepackage[margin=1in]{geometry}
\usepackage[utf8]{inputenc}
\usepackage{tikz}
\usepackage[compat=1.1.0]{tikz-feynman}
\usepackage[T1]{fontenc}
\usepackage{amssymb}
\usepackage{xcolor}
\usepackage{amsmath}
\usepackage{slashed}
\usepackage{braket}
\usepackage[normalem]{ulem}
\usepackage{comment}
\usepackage[caption=false]{subfig}
\usepackage{graphics}
\usepackage{graphicx}
\usepackage{adjustbox}
\usepackage{hyperref}
\hypersetup{colorlinks,linkcolor={blue},citecolor={blue},urlcolor={blue}} 

\begin{document}

\title{Prospects for detecting axionlike particles\\at the Coherent CAPTAIN-Mills experiment}

\preprint{LA-UR-21-28474}

\affiliation{Bartoszek~Engineering,~Aurora,~IL~60506,~USA}
\affiliation{Columbia~University,~New~York,~NY~10027,~USA}
\affiliation{University~of~Edinburgh,~Edinburgh,~United~Kingdom}
\affiliation{Embry$-$Riddle~Aeronautical~University,~Prescott,~AZ~86301,~USA }
\affiliation{University~of~Florida,~Gainesville,~FL~32611,~USA}
\affiliation{Los~Alamos~National~Laboratory,~Los~Alamos,~NM~87545,~USA}
\affiliation{Massachusetts~Institute~of~Technology,~Cambridge,~MA~02139,~USA}
\affiliation{Universidad~Nacional~Aut\'{o}noma~de~M\'{e}xico,~CDMX~04510,~M\'{e}xico}
\affiliation{University~of~New~Mexico,~Albuquerque,~NM~87131,~USA}
\affiliation{New~Mexico~State~University,~Las~Cruces,~NM~88003,~USA}
\affiliation{Texas~A$\&$M~University,~College~Station,~TX~77843,~USA}

\author{A.A.~Aguilar-Arevalo}
\affiliation{Universidad~Nacional~Aut\'{o}noma~de~M\'{e}xico,~CDMX~04510,~M\'{e}xico}
\author{D.~S.\,M.~Alves}
\affiliation{Los~Alamos~National~Laboratory,~Los~Alamos,~NM~87545,~USA}
\author{S.~Biedron}
\affiliation{University~of~New~Mexico,~Albuquerque,~NM~87131,~USA}
\author{J.~Boissevain}
\affiliation{Bartoszek~Engineering,~Aurora,~IL~60506,~USA}
\author{M.~Borrego}
\affiliation{Los~Alamos~National~Laboratory,~Los~Alamos,~NM~87545,~USA}
\author{L.~Bugel}
\affiliation{Massachusetts~Institute~of~Technology,~Cambridge,~MA~02139,~USA}
\author{M.~Chavez-Estrada}
\affiliation{Universidad~Nacional~Aut\'{o}noma~de~M\'{e}xico,~CDMX~04510,~M\'{e}xico}
\author{J.M.~Conrad}
\affiliation{Massachusetts~Institute~of~Technology,~Cambridge,~MA~02139,~USA}
\author{R.L.~Cooper}
\affiliation{Los~Alamos~National~Laboratory,~Los~Alamos,~NM~87545,~USA}
\affiliation{New~Mexico~State~University,~Las~Cruces,~NM~88003,~USA}
\author{A.~Diaz}
\affiliation{Massachusetts~Institute~of~Technology,~Cambridge,~MA~02139,~USA}
\author{J.R.~Distel}
\affiliation{Los~Alamos~National~Laboratory,~Los~Alamos,~NM~87545,~USA}
\author{J.C.~D’Olivo}
\affiliation{Universidad~Nacional~Aut\'{o}noma~de~M\'{e}xico,~CDMX~04510,~M\'{e}xico}
\author{E.~Dunton}
\affiliation{Columbia~University,~New~York,~NY~10027,~USA}
\author{B.~Dutta}
\affiliation{Texas~A$\&$M~University,~College~Station,~TX~77843,~USA}
\author{D.~Fields}
\affiliation{University~of~New~Mexico,~Albuquerque,~NM~87131,~USA}
\author{J.R.~Gochanour}
\affiliation{Los~Alamos~National~Laboratory,~Los~Alamos,~NM~87545,~USA}
\author{M.~Gold}
\affiliation{University~of~New~Mexico,~Albuquerque,~NM~87131,~USA}
\author{E.~Guardincerri}
\affiliation{Los~Alamos~National~Laboratory,~Los~Alamos,~NM~87545,~USA}
\author{E.C.~Huang}
\affiliation{Los~Alamos~National~Laboratory,~Los~Alamos,~NM~87545,~USA}
\author{N.~Kamp}
\affiliation{Massachusetts~Institute~of~Technology,~Cambridge,~MA~02139,~USA}
\author{D.~Kim}
\affiliation{Texas~A$\&$M~University,~College~Station,~TX~77843,~USA}
\author{K.~Knickerbocker}
\affiliation{Los~Alamos~National~Laboratory,~Los~Alamos,~NM~87545,~USA}
\author{W.C.~Louis}
\affiliation{Los~Alamos~National~Laboratory,~Los~Alamos,~NM~87545,~USA}
\author{J.T.M.~Lyles}
\affiliation{Los~Alamos~National~Laboratory,~Los~Alamos,~NM~87545,~USA}
\author{R.~Mahapatra}
\affiliation{Texas~A$\&$M~University,~College~Station,~TX~77843,~USA}
\author{S.~Maludze}
\affiliation{Texas~A$\&$M~University,~College~Station,~TX~77843,~USA}
\author{J.~Mirabal}
\affiliation{Los~Alamos~National~Laboratory,~Los~Alamos,~NM~87545,~USA}
\author{D.~Newmark}
\affiliation{Massachusetts~Institute~of~Technology,~Cambridge,~MA~02139,~USA}
\author{N.~Mishra}
\affiliation{Texas~A$\&$M~University,~College~Station,~TX~77843,~USA}
\author{P.~deNiverville}
\affiliation{Los~Alamos~National~Laboratory,~Los~Alamos,~NM~87545,~USA}
\author{V.~Pandey}
\affiliation{University~of~Florida,~Gainesville,~FL~32611,~USA}
\author{D.~Poulson}
\affiliation{Los~Alamos~National~Laboratory,~Los~Alamos,~NM~87545,~USA}
\author{H.~Ray}
\affiliation{University~of~Florida,~Gainesville,~FL~32611,~USA}
\author{E.~Renner}
\affiliation{Los~Alamos~National~Laboratory,~Los~Alamos,~NM~87545,~USA}
\author{T.J.~Schaub}
\affiliation{University~of~New~Mexico,~Albuquerque,~NM~87131,~USA}
\author{A.~Schneider}
\affiliation{Massachusetts~Institute~of~Technology,~Cambridge,~MA~02139,~USA}
\author{M.H.~Shaevitz}
\affiliation{Columbia~University,~New~York,~NY~10027,~USA}
\author{D.~Smith}
\affiliation{Embry$-$Riddle~Aeronautical~University,~Prescott,~AZ~86301,~USA }
\author{W.~Sondheim}
\affiliation{Los~Alamos~National~Laboratory,~Los~Alamos,~NM~87545,~USA}
\author{A.M.~Szelc}
\affiliation{University~of~Edinburgh,~Edinburgh,~United~Kingdom}
\author{C.~Taylor}
\affiliation{Los~Alamos~National~Laboratory,~Los~Alamos,~NM~87545,~USA}
\author{A.~Thompson}
\affiliation{Texas~A$\&$M~University,~College~Station,~TX~77843,~USA}
\author{W.H.~Thompson}
\affiliation{Los~Alamos~National~Laboratory,~Los~Alamos,~NM~87545,~USA}
\author{M.~Tripathi}
\affiliation{University~of~Florida,~Gainesville,~FL~32611,~USA}
\author{R.T.~Thornton}
\affiliation{Los~Alamos~National~Laboratory,~Los~Alamos,~NM~87545,~USA}
\author{R.~Van~Berg}
\affiliation{Bartoszek~Engineering,~Aurora,~IL~60506,~USA}
\author{R.G.~Van~de~Water}
\affiliation{Los~Alamos~National~Laboratory,~Los~Alamos,~NM~87545,~USA}
\author{S.~Verma}
\affiliation{Texas~A$\&$M~University,~College~Station,~TX~77843,~USA}

\collaboration{The CCM Collaboration}

\begin{abstract}
We show results from the Coherent CAPTAIN Mills (CCM) 2019 engineering run which begin to constrain regions of parameter space for axion-like particles (ALPs) produced in electromagnetic particle showers in an 800 MeV proton beam dump, and further investigate the sensitivity of ongoing data-taking campaigns for the CCM200 upgraded detector. Based on beam-on background estimates from the engineering run, we make realistic extrapolations for background reduction based on expected shielding improvements, reduced beam width, and analysis-based techniques for background rejection. We obtain reach projections for two classes of signatures; ALPs coupled primarily to photons can be produced in the tungsten target via the Primakoff process, and then produce a gamma-ray signal in the Liquid Argon (LAr) CCM detector either via inverse Primakoff scattering or decay to a photon pair. ALPs with significant electron couplings have several additional production mechanisms (Compton scattering, $e^+e^-$ annihilation, ALP-bremsstrahlung) and detection modes (inverse Compton scattering, external $e^+e^-$ pair conversion, and decay to $e^+e^-$). In some regions, the constraint is marginally better than  both astrophysical and terrestrial constraints. With the beginning of a three year run, CCM will be more sensitive to this parameter space by up to an order of magnitude for both ALP-photon and ALP-electron couplings. The CCM experiment will also have sensitivity to well-motivated parameter space of QCD axion models. It is only a recent realization that accelerator-based large volume liquid argon detectors designed for low energy coherent neutrino and dark matter scattering searches are also ideal for probing ALPs in the unexplored $\sim$MeV mass scale.
\end{abstract}

\maketitle

\section{Introduction}
Axion-like particles (ALPs) are generically predicted in a variety of well-motivated beyond the Standard Model (BSM) scenarios, including the Peccei-Quinn mechanism that solves the Strong CP problem~\cite{Peccei:1977hh} (which predicts the QCD axion~\cite{Wilczek:1977pj,Weinberg:1977ma}), and dark sectors encompassing dark matter, dark mediators, and BSM neutrino physics~\cite{Lanfranchi:2020crw}.
The vast ALP parameter space, which spans many orders of magnitude in mass and couplings to photons, electrons, and nucleons, is being probed by a broad experimental effort. These include ongoing and proposed experiments such as haloscopes (ADMX~\cite{Asztalos:2001tf,Du:2018uak}, ABRACADABRA~\cite{Kahn:2016aff,Salemi:2019xgl}, HAYSTAC~\cite{Brubaker:2016ktl,Droster:2019fur}, CASPEr~\cite{JacksonKimball:2017elr}), helioscopes (CAST~\cite{Zioutas:1998cc,Anastassopoulos:2017ftl}, IAXO \cite{Irastorza:2013dav,IAXO:2019mpb}), interferometry \cite{Melissinos:2008vn,DeRocco:2018jwe,Obata:2018vvr,Liu:2018icu}, light-shining-through-wall experiments \cite{Spector:2019ooq},  ongoing and future accelerator-based experiments (NA62 \cite{Volpe:2019nzt}, NA64~\cite{Dusaev:2020gxi,Banerjee:2020fue}, FASER \cite{Feng:2018noy}, LDMX \cite{Berlin:2018bsc,Akesson:2018vlm},  SeaQuest~\cite{Berlin:2018pwi}, SHiP \cite{Alekhin:2015byh}, PASSAT~\cite{Bonivento:2019sri}),
reactor experiments (e.g., MINER, CONUS, TEXONO etc.~\cite{Dent:2019ueq,AristizabalSierra:2020rom,Chang:2006ug}), dark matter experiments (DAMA~\cite{Bernabei:2001ny}, XMASS \cite{Oka:2017rnn}, EDELWEISS \cite{Armengaud:2013rta,Armengaud:2018cuy},  SuperCDMS~\cite{PhysRevD.101.052008}, XENON~\cite{Aprile:2020tmw,Dent:2020jhf},  PandaX~\cite{PandaX:2017ock}), resonant absorption by nuclei~\cite{Moriyama:1995bz,Krcmar:1998xn,Krcmar:2001si,Derbin:2009jw,Gavrilyuk:2018jdi,Creswick:2018stb,Li:2015tsa,Li:2015tyq,Benato:2018ijc,Dent:2021jnf}, astrophysical observations~\cite{Dent:2020qev, Carenza:2021alz, Galanti:2018nvl,Tavecchio:2012um,Galanti:2015rda,Ayala:2014pea,Fermi-LAT:2016nkz,Conlon:2013txa,Conlon:2015uwa,Conlon:2017qcw,Raffelt:1987im}, etc.

In the 70s and 80s, the QCD axion was extensively searched for in beam dump, fixed target, and reactor experiments (see, for example, refs.~\cite{Donnelly:1978ty, Bjorken:1988as, Chang:2006ug, PhysRevLett.59.755, Lehmann:1982bp, Avignone:1988bv, Bechis:1979kp, Riordan:1987aw, Bross:1989mp}). This effort has been revived with the modern Intensity Frontier experimental program to explore more generic ALP signals. Over the past three decades advances in accelerators have enabled modern experiments to gain an order of magnitude or more in instantaneous beam luminosity, while advances in instrumentation have led to significant improvements in detection efficiency, in energy, spatial, and timing resolution, lower detection thresholds, and particle identification (see e.g., refs.~\cite{Klein:2022lrf, Batell:2022xau, Huber:2022lpm, Ilten:2022lfq} for a review). In particular, the high intensity photon flux and associated electromagnetic cascades in reactor and accelerator neutrino experiments offer new opportunities to explore ALP production via its electromagnetic and leptonic couplings~\cite{Dent:2019ueq, AristizabalSierra:2020rom, Brdar:2020dpr}.

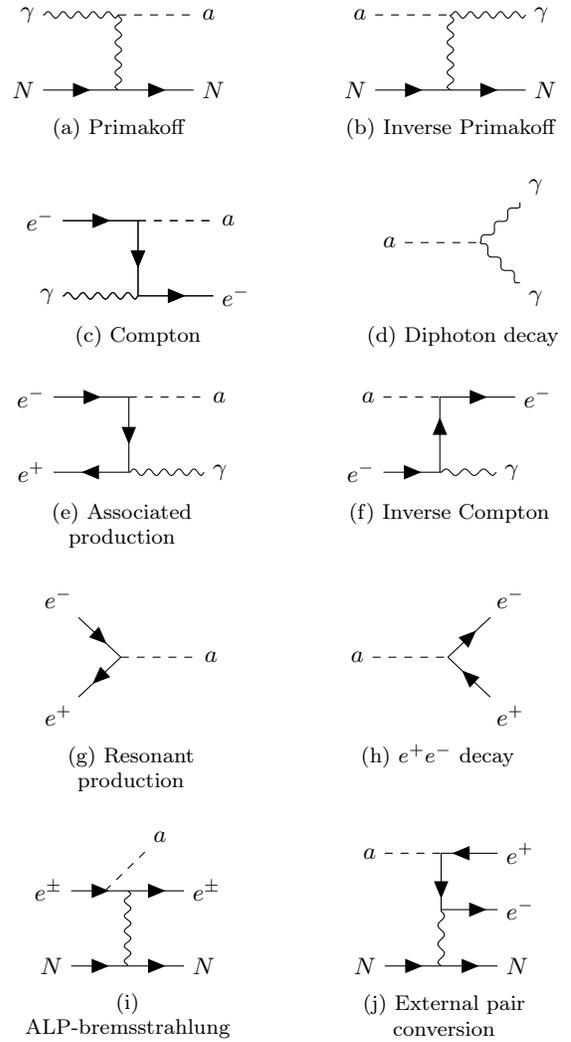
\begin{figure}[tbh]
 \centering
\subfloat[Primakoff]{
     \begin{tikzpicture}
              \begin{feynman}
         \vertex (o1);
         \vertex [right=1cm of o1] (f1) {\(a\)};
         \vertex [left=1cm of o1] (i1){\(\gamma\)} ;
         \vertex [below=1cm of o1] (o2);
         \vertex [right=1cm of o2] (f2) {\(N\)};
         \vertex [left=1cm of o2] (i2) {\(N\)};

         \diagram* {
           (i1) -- [boson] (o1) -- [scalar] (f1),
           (o1) -- [boson] (o2),
           (i2) -- [fermion] (o2),
           (o2) -- [ fermion] (f2),
         };
        \end{feynman} 
       \end{tikzpicture}
       \label{fig:axionPrimakoff}
}\hspace{30pt}
\subfloat[Inverse Primakoff]{  
       \begin{tikzpicture}
              \begin{feynman}
         \vertex (o1);
         \vertex [right=1cm of o1] (f1) {\(\gamma\)};
         \vertex [left=1cm of o1] (i1){\(a\)} ;
         \vertex [below=1cm of o1] (o2);
         \vertex [right=1cm of o2] (f2) {\(N\)};
         \vertex [left=1cm of o2] (i2) {\(N\)};

         \diagram* {
           (i1) -- [scalar] (o1) -- [boson] (f1),
           (o1) -- [boson] (o2),
           (i2) -- [fermion] (o2),
           (o2) -- [fermion] (f2),
         };
        \end{feynman}
       \end{tikzpicture}
       \label{fig:axionInvPrimakoff}
} \\
\subfloat[Compton]{
    \begin{tikzpicture}
        \begin{feynman}
         
        \vertex (o1);
         \vertex [right=1cm of o1] (f1) {\(a\)};
         \vertex [left=1cm of o1] (i1){\(e^-\)} ;
         \vertex [below=1cm of o1] (o2);
         \vertex [right=1cm of o2] (f2) {\(e^-\)};
         \vertex [left=1cm of o2] (i2) {\(\gamma\)};

         \diagram* {
           (i1) -- [fermion] (o1) -- [scalar] (f1),
           (o1) -- [fermion] (o2),
           (i2) -- [boson] (o2),
           (o2) -- [ fermion] (f2),
         };
         
          \diagram* {
           (i1) -- [fermion] (o1) -- [scalar] (f1),
           (o1) -- [fermion] (o2),
           (i2) -- [boson] (o2),
           (o2) -- [ fermion] (f2),
         };
        \end{feynman}
    \end{tikzpicture}
    \label{fig:axionCompton}
}\hspace{35pt}
\subfloat[Diphoton decay]{
       \begin{tikzpicture}
       \begin{feynman}
         \vertex (o1);
         \vertex [left=1cm of o1] (i) {\(a\)};
         \vertex [above right=0.75cm of o1] (f1) {\(\gamma\)};
         \vertex [below right=0.75cm of o1] (f2) {\(\gamma\)};

         \diagram* {
           (i) -- [scalar] (o1),
           (o1) -- [boson] (f1),
           (o1) -- [boson] (f2),
         };
        \end{feynman}
       \end{tikzpicture}
        \label{fig:axionDecayDiphoton}
} \\
\subfloat[Associated production]{
    \begin{tikzpicture}
              \begin{feynman}
         \vertex (o1);
         \vertex [left=1.0cm of o1] (i1) {\(e^-\)};
         \vertex [right=1.0cm of o1] (f1){\(a\)};
         \vertex [below=1.0cm of o1] (o2);
         \vertex [right=1.0cm of o2] (f2){\(\gamma\)};
         \vertex [left=1.0cm of o2] (i2) {\(e^+\)};

         \diagram* {
           (i1) -- [fermion] (o1) -- [fermion] (o2) -- [fermion] (i2),
           (o2) -- [boson] (f2),
           (o1) -- [scalar] (f1),
         };
        \end{feynman}
       \end{tikzpicture}
    \label{fig:axionAssociated}
}\hspace{30pt}
\subfloat[Inverse Compton]{
    \begin{tikzpicture}
        \begin{feynman}
         \vertex (o1);
         \vertex [right=1.0cm of o1] (f1) {\(e^-\)};
         \vertex [left=0.75 of o1] (i1){\(a\)} ;
         \vertex [below=1cm of o1] (o2);
         \vertex [right=0.75 of o2] (f2) {\(\gamma\)};
         \vertex [left=0.75 of o2] (i2) {\(e^-\)};

         \diagram* {
           (i1) -- [scalar] (o1) -- [fermion] (f1),
           (o1) -- [anti fermion] (o2),
           (i2) -- [fermion] (o2),
           (o2) -- [ boson] (f2),
         };
        \end{feynman}
    \end{tikzpicture}
    \label{fig:axionInvCompton}
} \\
\subfloat[Resonant production]{
    \begin{tikzpicture}
              \begin{feynman}
         \vertex (o1);
         \vertex [above left=0.75cm of o1] (i1) {\(e^-\)};
         \vertex [below left=0.75cm of o1] (i2) {\(e^+\)};
         \vertex [right=1.0cm of o1] (f1) {\(a\)};
         \diagram* {
           (i1) -- [fermion] (o1) -- [fermion] (i2),
           (o1) -- [scalar] (f1)
         };
        \end{feynman}
       \end{tikzpicture}
    \label{fig:axionProductionResonance}
}\hspace{35pt}
\subfloat[$e^+e^-$ decay]{
    \begin{tikzpicture}\begin{feynman}
         \vertex (o1);
         \vertex [left=1cm of o1] (i) {\(a\)};
         \vertex [above right=0.75cm of o1] (f1) {\(e^-\)};
         \vertex [below right=0.75cm of o1] (f2) {\(e^+\)};

         \diagram* {
           (i) -- [scalar] (o1),
           (f2) -- [fermion] (o1) -- [fermion] (f1),
         };
        \end{feynman}
    \end{tikzpicture}
    \label{fig:axionDecayElectronPositron}
} \\
\subfloat[ALP-bremsstrahlung]{
    \begin{tikzpicture}
              \begin{feynman}
         \vertex (o1);
         \vertex [left=0.75cm of o1] (i1) {\(e^\pm\)};
         \vertex [right=0.75cm of o1] (f1) {\(e^\pm\)};
         \vertex [below=1cm of o1] (o2);
         \vertex [right=0.75cm of o2] (f2) {\(N\)};
         \vertex [left=0.75cm of o2] (i2) {\(N\)};
         \vertex [left=0.3cm of o1] (b1);
         \vertex [above right=0.75cm of b1] (f3) {\(a\)};

         \diagram* {
           (i1) -- [fermion] (o1) -- [fermion] (f1),
           (i2) -- [fermion] (o2) -- [fermion] (f2),
           (o2) -- [boson] (o1),
           (b1) -- [scalar] (f3),
         };
        \end{feynman}
       \end{tikzpicture}
    \label{fig:axionBrem}
}\hspace{35pt}
\subfloat[External pair conversion]{
    \begin{tikzpicture}
              \begin{feynman}
         \vertex (o1);
         \vertex [left=0.75cm of o1] (i1) {\(a\)};
         \vertex [right=0.75cm of o1] (f1){\(e^+\)};
         \vertex [below=0.75cm of o1] (o2);
         \vertex [right=0.75cm of o2] (f2){\(e^-\)};
         \vertex [below=0.75cm of o2] (o3);
         \vertex [left=0.75cm of o3] (i2) {\(N\)};
         \vertex [right=0.75cm of o3] (f3) {\(N\)};

         \diagram* {
           (i1) -- [scalar] (o1),
           (f2) -- [anti fermion] (o2) -- [anti fermion] (o1) -- [anti fermion] (f1),
           (o2) -- [boson] (o3),
           (i2) -- [fermion] (o3),
           (o3) -- [fermion] (f3),
         };
        \end{feynman}
       \end{tikzpicture}
    \label{fig:axionPairProduction}
}
    \caption{Processes contributing to ALP production (left column) and detection (right column) considered in this analysis.}
    \label{fig:axion}
\end{figure}

For ALPs that couple predominantly to photons, the Primakoff and inverse-Primakoff processes can be exploited for ALP production and detection, respectively. These processes, illustrated in Figs.\,\ref{fig:axionPrimakoff}, \ref{fig:axionInvPrimakoff}, are coherently enhanced by a factor of $Z^2$, where $Z$ is the atomic number of the target nucleus. The ALP flux can produce electromagnetic signals in the detector via inverse Primakoff scattering or decay to a photon pair within the detector's fiducial volume (Fig.\,\ref{fig:axionDecayDiphoton}).

For ALPs with significant couplings to electrons, other processes can also contribute to its production, including Compton-like scattering (Fig.\,\ref{fig:axionCompton}), $e^+e^-$ annihilation (Figs.\,\ref{fig:axionAssociated}, \ref{fig:axionProductionResonance}), and ALP-bremsstrahlung (Fig.\,\ref{fig:axionBrem}). Such ALPs can be detected via inverse-Compton scattering (Fig.\,\ref{fig:axionInvCompton}) and $e^+e^-$ conversion (Fig.\,\ref{fig:axionPairProduction}), or, if sufficiently short-lived, via decay to $e^+e^-$ within the detector (Fig.\,\ref{fig:axionDecayElectronPositron}). Detailed descriptions of the amplitudes relevant for these processes are given in Appendix~\ref{app:cross_sections}.

\begin{figure*}[!hbt]
\centering
    \includegraphics[width=0.8\textwidth]{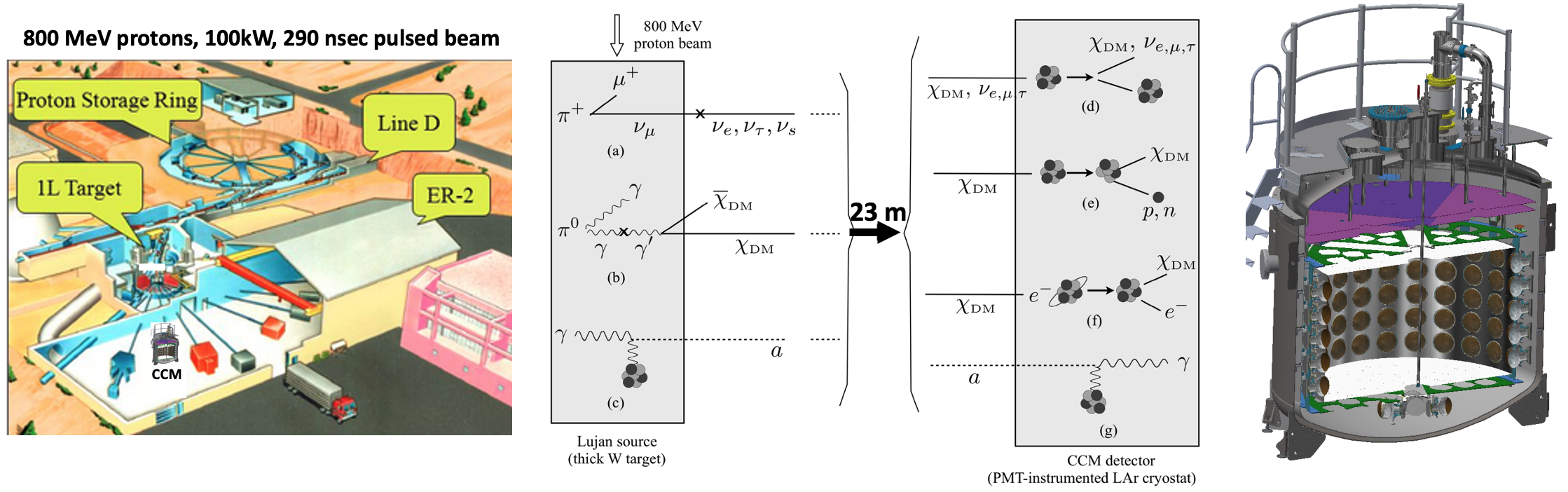}
    \caption{CCM experiment layout. On the left protons from the LANSCE accelerator are compressed in the Proton Storage Ring (PSR) to pulses of 290\,ns width at 20\,Hz.  They impact the tungsten target from above making all experiments on the Lujan floor 90$^\circ$ from the beam axis.  The CCM detector (right) is placed 23\,m away from the target.  There is approximately 5\,m of steel, 2\,m concrete, and 10\,cm borated polly shielding between the target and detector to reduce fast and thermal neutrons.  The middle figure shows some of the various production and detection processes occurring in the experiment.}
    \label{fig:CCM}
\end{figure*}

Previous studies have shown that ongoing reactor-based neutrino experiments such as CONUS, CONNIE, MINER etc.~\cite{Dent:2019ueq,AristizabalSierra:2020rom}, and upcoming accelerator-based neutrino experiments such as DUNE~\cite{Brdar:2020dpr} will be able to probe parameter space for ALPs in the MeV mass range coupling to electrons, photons, and nucleons. This mass range has remained inaccessible to terrestrial and astrophysical observations to date. In this paper, we investigate the sensitivity of the Coherent CAPTAIN-Mills (CCM) experiment to ALPs coupled electromagnetically or electronically and push the sensitivity envelope in the MeV mass region of ALP parameter space. The LANSCE accelerator and proton storage ring provides an 800 MeV, 100\,$\mu$A, short 290\, ns pulse of protons (triangular shape) impinging on a thick tungsten target and producing significant hadronic activity and a high intensity flux of photons and electromagnetic cascades in the $\mathcal{O}(0.1-1000)$ MeV energy range. ALPs produced through photons and $e^\pm$ interacting with the tungsten target material could be detected at the 5-ton (fiducial cylinder approximately 1\,m height by 2\,m diameter) liquid Argon detector located 23 meters away from the target and 90$^\circ$ from the beam direction, as shown in Fig.\,\ref{fig:CCM}.  ALP's that interact electromagnetically in the CCM liquid Argon will produce a copious shower of scintillation light ($\sim$40,000 photons/MeV) at 128\,nm with a prompt 6\,nsec and slower 1.6\,$\mu$sec decay constant~\cite{PhysRevB.27.5279, LArproperties}.  The scintillation light is wavelength shifted by Tetraphenyl Butadiene (TPB) surfaces and then detected by an array of PMT's. See \cite{CCM:2021leg} for detection details.

In 2019 a six week engineering beam run was performed with the CCM120 detector, named due to it having 120 inward pointing main PMTs. The CCM120 experiment met expectations and performed a sensitive search for sub-GeV dark matter via coherent nuclear scattering with $1.79\times10^{21}$ Protons On Target (POT)~\cite{CCM:2021leg,CCM:2021yzc}. Due to the intense scintillation light production and short 14\,cm radiation length in LAr \cite{LArproperties}, the relatively large CCM detector has good response to electromagnetic signal events in the energy range from 100 keV up to 10's of MeV.  This low energy kinematic range, which is sensitive to $\sim$MeV ALP mass range,  has not been previously explored at proton beam dump experiments.  CCM's novel sensitivity to this region could probe new parameter space of BSM particle production and yield new insights into the nature of the LSND \cite{LSND:2001aii} and MiniBooNE event excesses \cite{MiniBooNE:2020pnu,Dutta:2021cip}. Another key feature of CCM is that it uses fast beam and detector timing to isolate prompt ultra-relativistic particles, which ALPs in the MeV mass range may be for the energy scale of the Lujan proton beam source. This can distinguish them from the significantly slower neutron backgrounds that arrive approximately 225\,ns after the start of the beam pulse (relativistic particles traverse the 23\,m distance in 76.6\,ns) \cite{CCM:2021leg}. Furthermore, the Lujan beam low duty factor of $\sim 10^{-5}$ and extensive shielding are efficient at rejecting steady state backgrounds from cosmic rays, neutron activation, and internal radioactivity from PMTs and $^{39}$Ar.   In order to determine the sensitivity reach of CCM's ongoing run, we use the  beam-on background distribution determined from the recent CCM120 run \cite{CCM:2021leg}, with a further expected factor of 100 reduction from extensive improvements in shielding, veto rejection, energy and spatial resolution, particle identification analysis, and reduced beam width.

In \S~\ref{sec:alp_model} we discuss the ALP phenomenological models, and their production and detection modes at CCM; in \S~\ref{sec:ccm120} we review the treatment of backgrounds and cuts to optimize the signal efficiency at CCM120; in \S~\ref{sec:reach} we set limits on the ALP parameter space from current CCM120 data and projected sensitivities for the ongoing 3-year physics run, and in \S~\ref{sec:conclusion} we conclude. Additional details on the signal prediction are outlined in Appendix~\ref{app:cross_sections} and details on the optical model and reconstruction in Appendix~\ref{app:smearing}.

\section{\bf{ALP} Models and Their Energy Spectr	a at CCM}\label{sec:alp_model}
The focus of this analysis is on generic models of ALPs with couplings to photons and electrons. These interactions are parameterized by the following Lagrangian terms:
\begin{equation}\label{ALPlagrangian}
\mathcal{L}_{\rm ALP} ~\supset~ -\frac{g_{a\gamma}}{4}\,a\,F_{\mu\nu}\tilde{F}^{\mu\nu}\,-\,g_{ae}\,a\,\bar e \,i \gamma_5\, e\,,
\end{equation}
where $e$ is the electron Dirac fermion, and ($\tilde{F}^{\mu\nu}$) $F_{\mu\nu}$ is the electromagnetic (dual-)field strength tensor.
For specific ALP models, including the QCD axion and its variants, the couplings $g_{a\gamma}$ and $g_{a e}$ are not independent. For the purposes of this study, however, we will adopt a simplified model approach by considering two limiting cases: in the first, we set $g_{a e}=0$, so that the ALP phenomenology is completely determined by its electromagnetic interactions parameterized by $g_{a\gamma}$; and in the second case, we assume that $g_{ae}$ is sufficiently large to dominate ALP production and detection at CCM---this limit holds when $g_{ae}\gg \alpha\, m_e g_{a \gamma}$.

\begin{figure}[h]
\centering
    \includegraphics[width=0.49\textwidth]{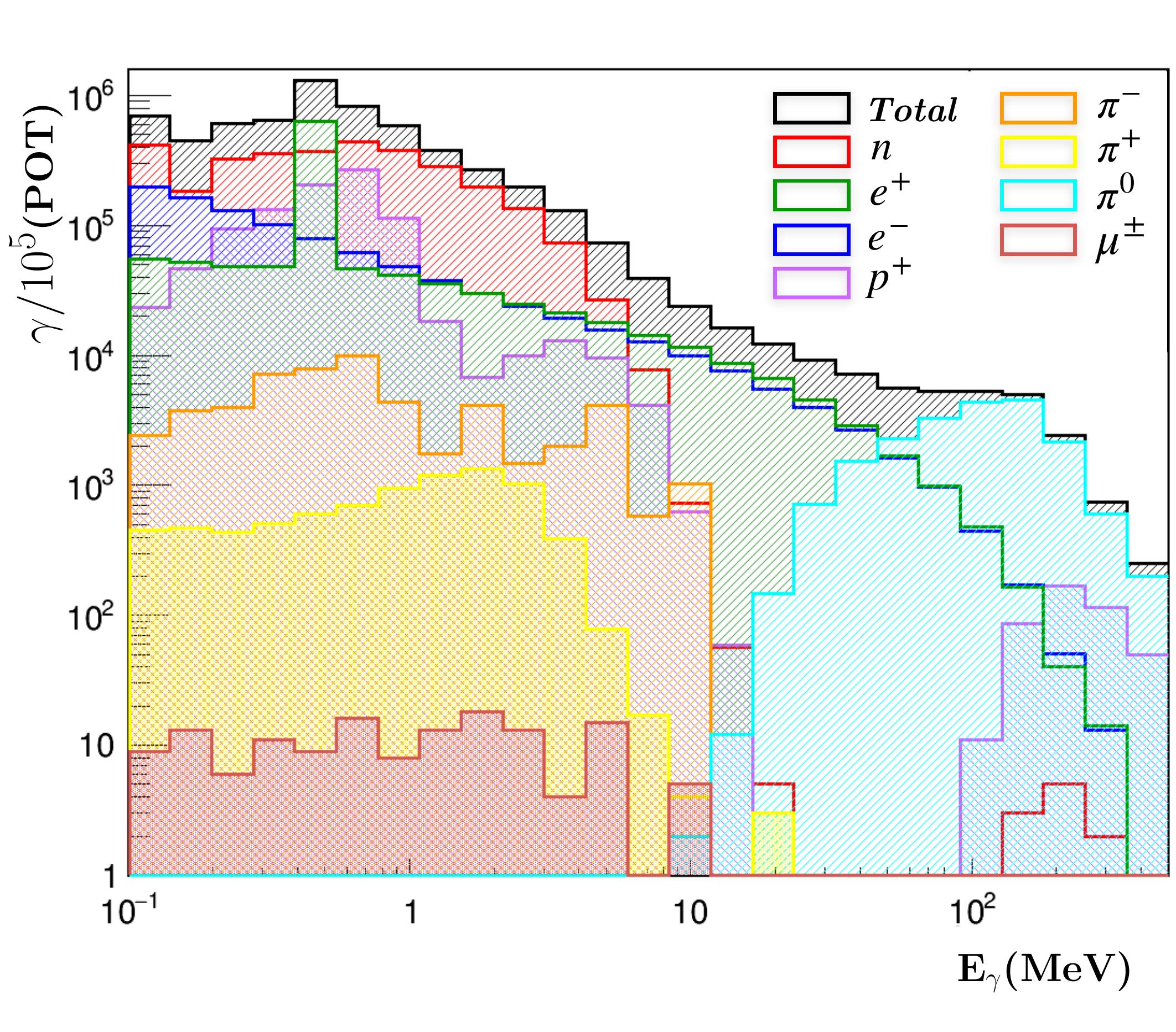}
    \includegraphics[width=0.49\textwidth]{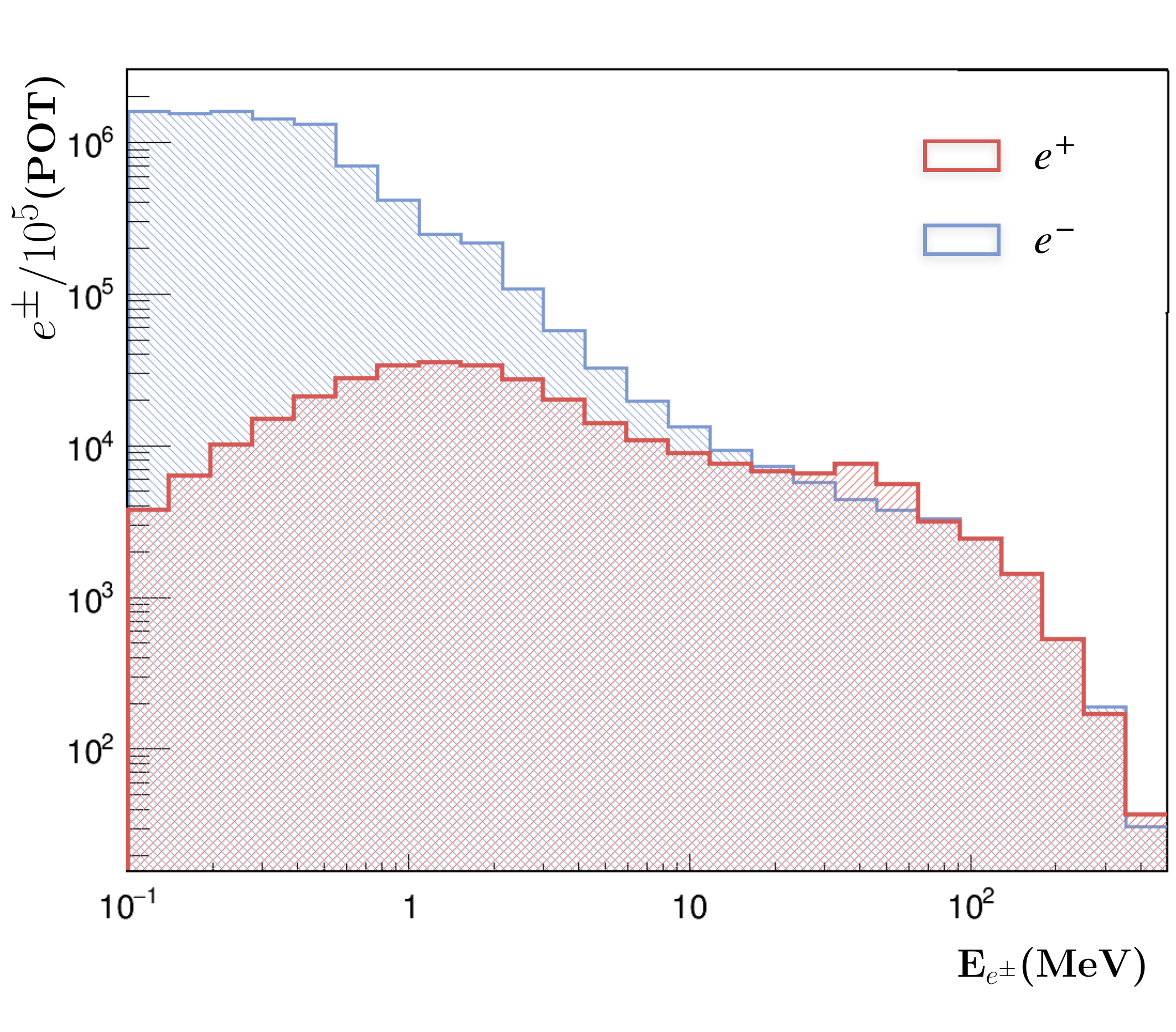}
    \caption{MC energy spectra for photons (top plot) and $e^\pm$ (bottom plot) at the Lujan source, simulated with \texttt{GEANT4 10.7} using the \texttt{QGSP\_BIC\_HP} library~\cite{Agostinelli:2002hh} by generating $10^5$ protons incident on a tungsten target.
    The different photoproduction sources are shown as non-stacked histograms in the top plot, with the total rate shown in black.}
    \label{fig:photonspec}
\end{figure}

At LANSCE's Lujan source, the 800 MeV proton beam impinging on the tungsten target produces a high intensity photon flux from cascades, neutron capture, pion decays, etc. These processes were modelled using \texttt{GEANT4 10.7} with the \texttt{QGSP\_BIC\_HP} library~\cite{Agostinelli:2002hh}, and the resulting photon, electron, and positron spectra are shown in Fig.\,\ref{fig:photonspec} for the energy range $E_{\gamma,e^\pm} = 0.1-500\, \text{MeV}$. Flux errors are primarily associated with differential neutron production uncertainties and for tungsten estimated to be less than 10\%~\cite{neutron-xs}. ALPs produced from the processes shown in Fig.\,\ref{fig:axion} could then propagate to the CCM detector, where they could produce an electromagnetic signal via inverse Primakoff or Compton-like scattering, diphoton decay, or $e^+e^-$ conversion or decay.

\begin{figure}[h]
\centering
    \includegraphics[width=0.49\textwidth]{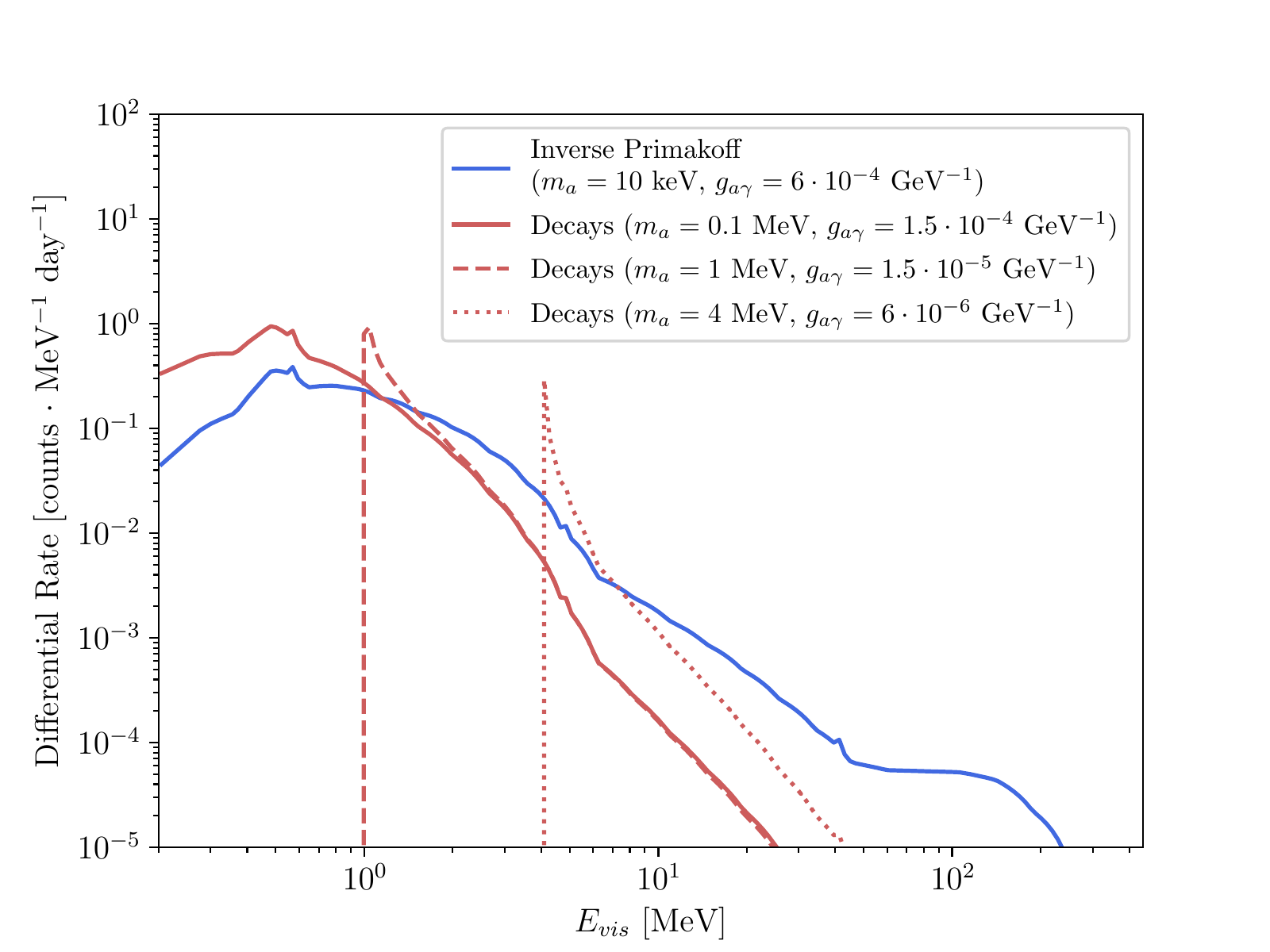}
    \includegraphics[width=0.49\textwidth]{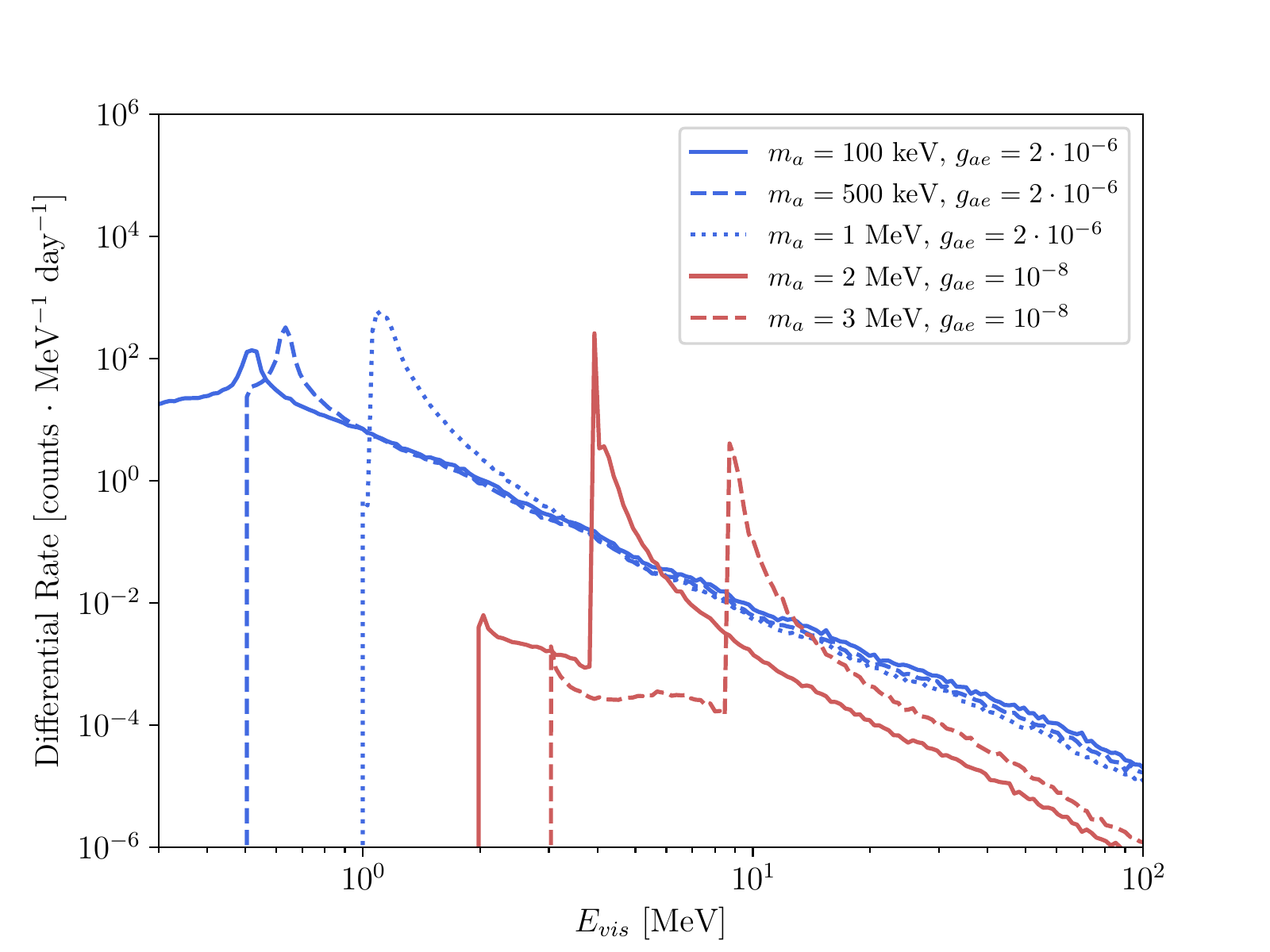}
    \caption{ALP-induced differential event rates at the CCM detector. Top panel: energy spectrum of 1-$\gamma$ and 2-$\gamma$ final states from inverse Primakoff scattering and diphoton decays. Bottom panel: energy spectrum of ($e^-$, $\gamma$) final states from inverse Compton scattering and $e^+e^-$ conversion/decays. In some scenarios the signal abruptly turns on, making it clearly identifiable relative to the backgrounds that generally fall off at higher energy (as described in \S.~\ref{sec:ccm120}).}
    \label{fig:spectra}
\end{figure}

 We show in the next section that the background spectrum relevant to these signal channels falls off exponentially at energies greater than a few MeV; thus, the ALP signals could be potentially visible due to the harder spectral shapes of $\gamma$'s and $e^\pm$'s from ALP scattering and decays. Monte Carlo simulations of the CCM detector response to gammas and electrons from ALP events indicate that the signal reconstruction efficiency for 5 fiducial tons of LAr is above 50\% for events above 100 keV.

In Fig.~\ref{fig:spectra} (top panel) we show the energy spectra of ALP-induced events with one or two photons in the final state resulting from either inverse Primakoff scattering or diphoton decays. Signals from heavier ALPs are characterized by a significantly harder spectrum, which in principle makes them more easily distinguishable from backgrounds. On the other hand, the ALP lifetime decreases with ALP mass at fixed coupling, and therefore heavier ALPs need to be more boosted in order to decay within the detector fiducial volume.

In Fig.\,\ref{fig:spectra} (bottom panel) we show the energy spectra of ALP-induced events with an electron or an $e^+e^-$ pair in the final state, arbitrarily normalized to yield a non-negligible signal-to-background ratio. At lower ALP masses $m_a\,\lesssim\, 2\,m_e$ the signal is dominated by inverse Compton scattering, whereas for ALP masses above the $e^+e^-$ threshold, the signal is dominated by $a\to e^+e^-$ decays exhibiting a ``line-shape'' spectral feature. This is because the dominant ALP production mechanism in this kinematic range is resonant $e^+e^-$ annihilation to an ALP of energy $E_a=m_a^2 / (2m_e)$. If the ALP lifetime is such that it decays outside the detector, the ALP signal is instead dominated by external ALP conversion to an $e^+e^-$ pair in the detector.

\section{ALP Searches at CCM120}\label{sec:ccm120}

\begin{figure}[h]
    \centering
    \includegraphics[width=0.49\textwidth]{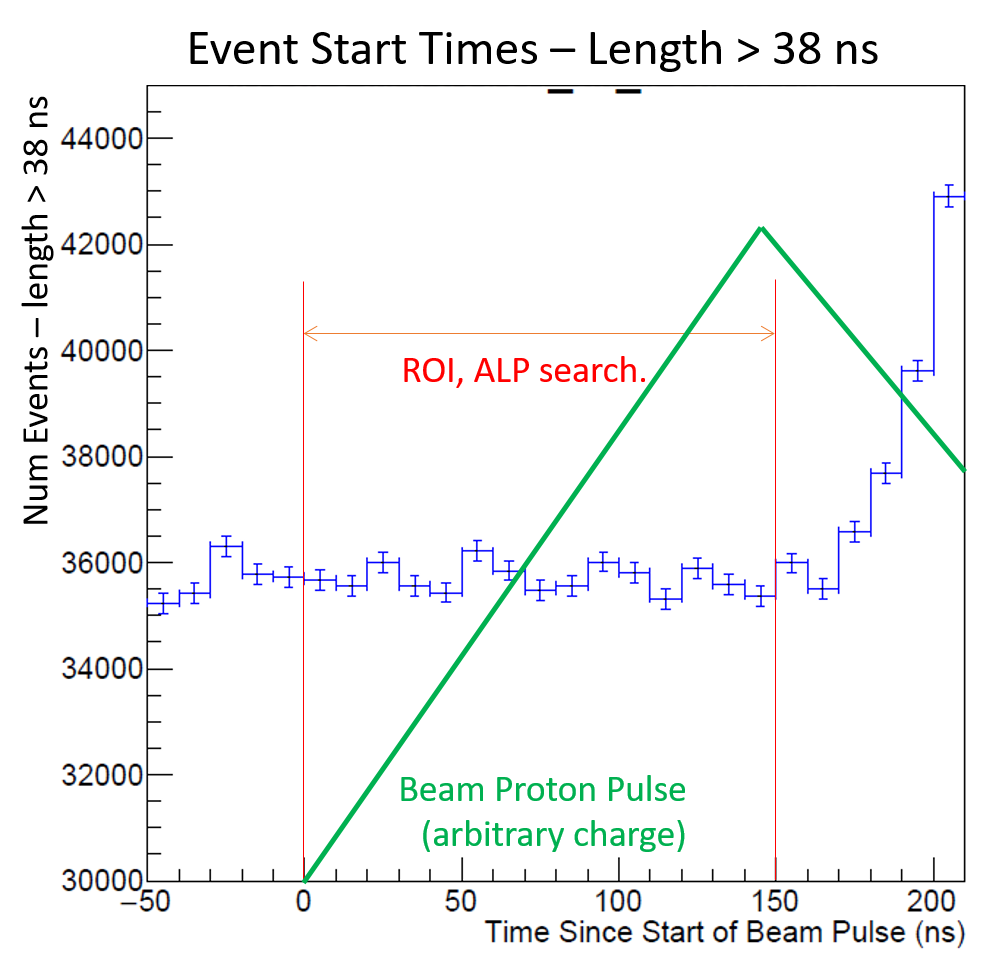}
    \caption{The event rate after the length cut in CCM120 around the measured beam T$_0$ (defined as T=0ns) when the earliest speed of light particles from the beam would arrive in CCM (see \cite{CCM:2021leg}).  For short nucleon scattering like events (<44ns)  the allowed ROI end time is 190ns after T$_0$. For long electromagnetic like events, changes in length cut efficiency began 170ns after T$_0$. For some energy bins efficiency changes began 150 ns after T$_0$, which defined the final analysis ROI of 150 ns.
    }
    \label{fig:roiEnd}
\end{figure}

\begin{figure}[h]
    \centering
    \includegraphics[width=0.4\textwidth]{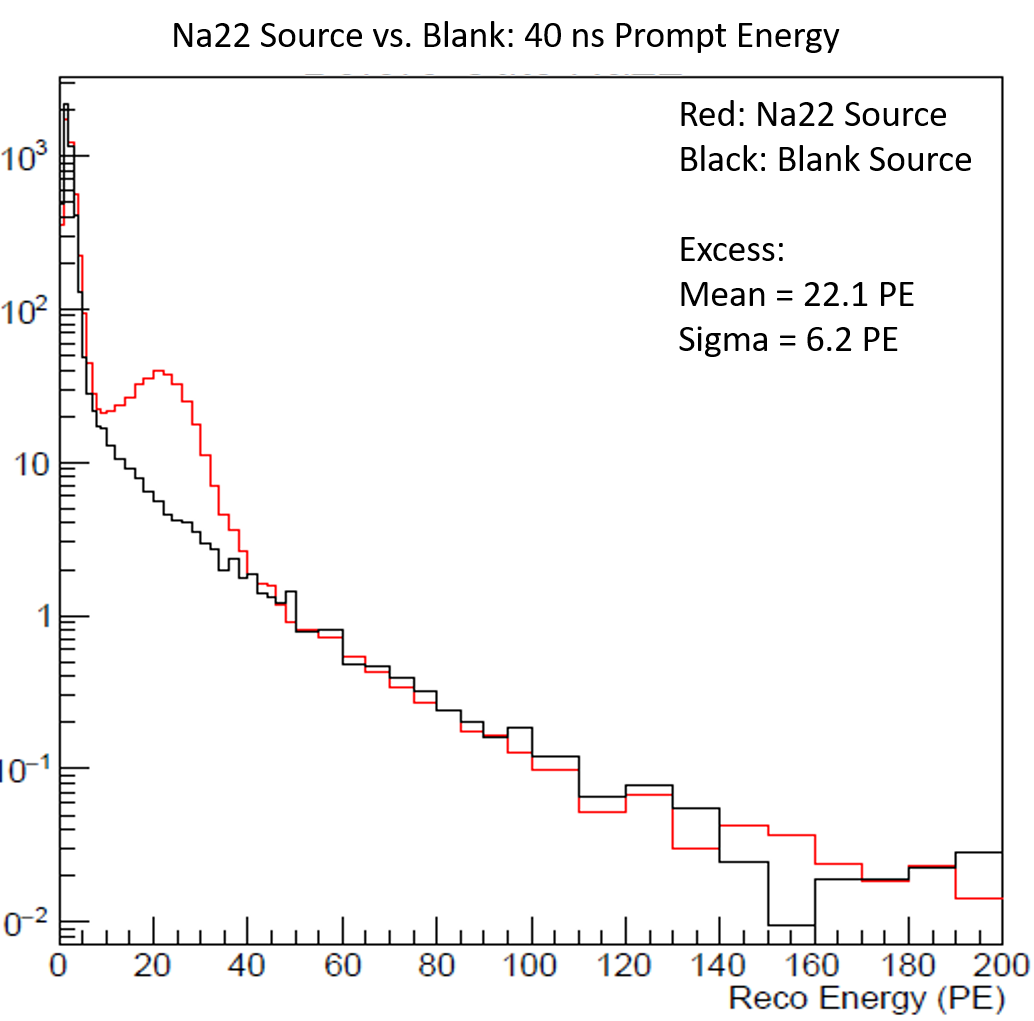}
    \caption{The $^{22}${Na} calibration used to determine the energy scale for the 40 ns prompt time. 40 ns of light accumulation was used to define event reconstructed energy in order to eliminate contamination from beam related neutrons coming in after the ROI.}
    \label{fig:40nsCalibration}
\end{figure}

\begin{figure}[h]
    \centering
    \includegraphics[width=0.4\textwidth]{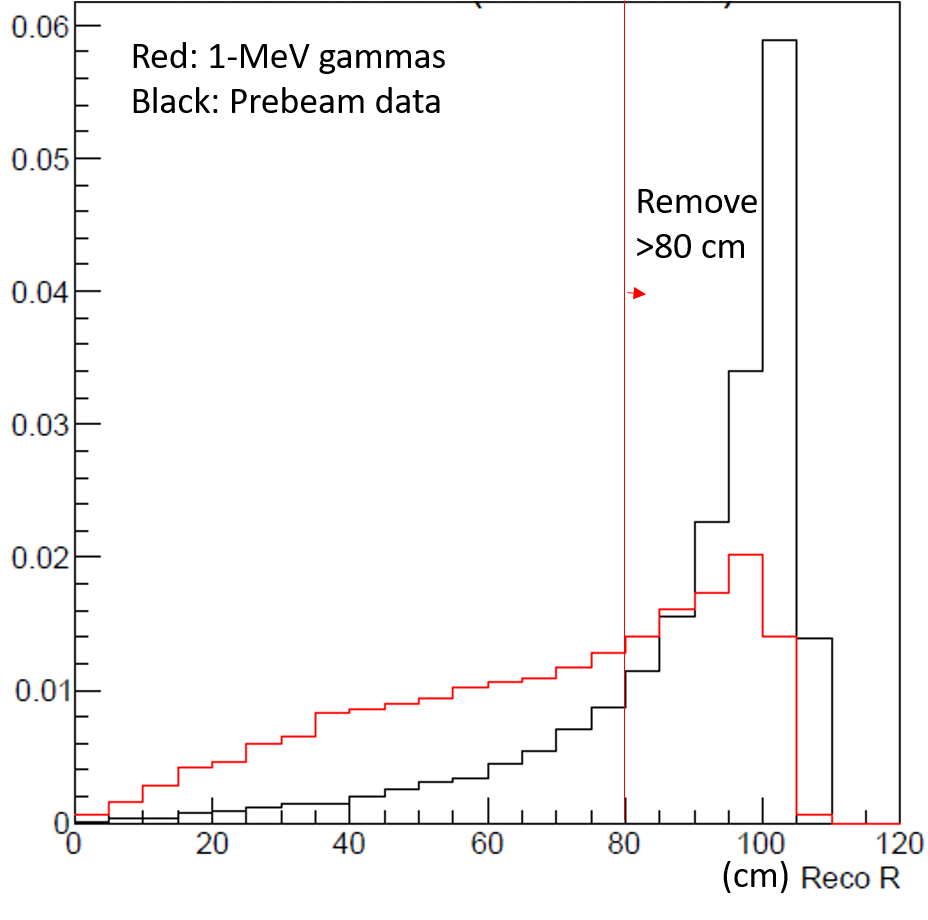}
    \caption{The ALP simulation (see Appendix~\ref{app:smearing}) vs. prebeam radius showing the preference for a tighter radius cut. ALP events followed an isotropic pattern throughout the detector, while background events from radioactivity preferentially occurred near the edges. The position resolution is $\pm10$ cm near the center, measured from calibration data.}
    \label{fig:alpRadius}
\end{figure}

\begin{figure}[h]
    \centering
    \includegraphics[width=0.4\textwidth]{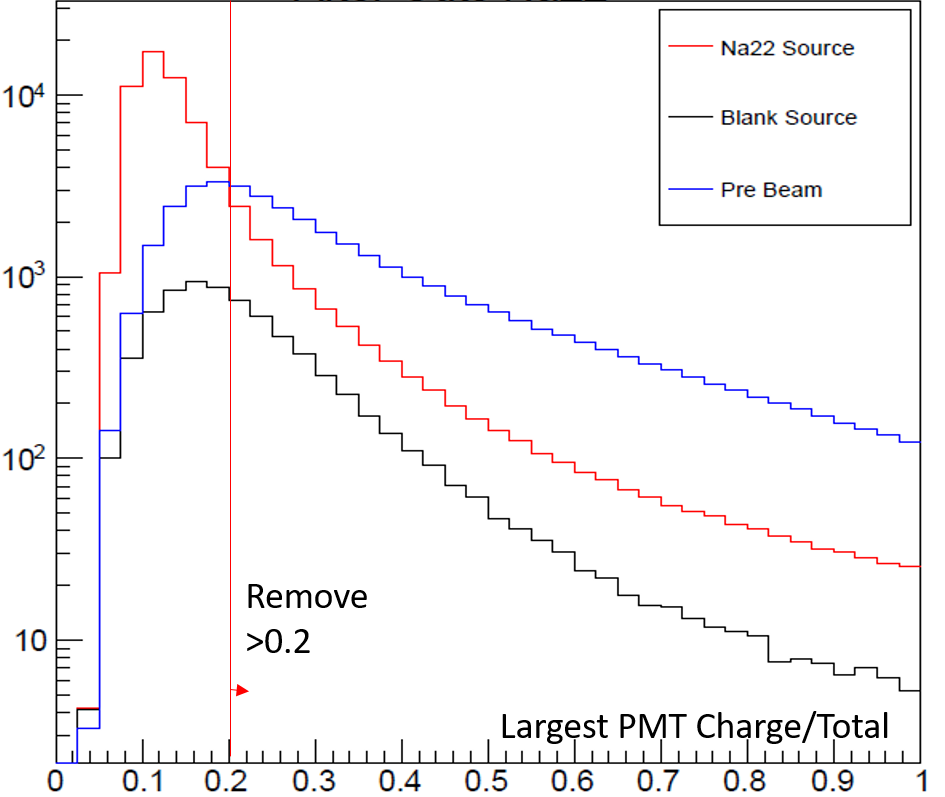}
    \caption{The shape difference between gamma-like $^{22}${Na} events and the overall prebeam in the maxPMT proportion variable. A figure of merit analysis placed a maximum value cut at 0.2.}
    \label{fig:maxPMT}
\end{figure}

In 2019 CCM120 carried out an initial engineering run to test the detector's capability to search for BSM physics, including ALPs \cite{thesis_Edward}. This search was composed of three steps: defining a beam-related neutron background free region of interest (ROI) in which to search for near speed of light ALPs, determining the signal characteristics of ALP events through analysis cuts to obtain the data, and defining confidence limits from the observed data excess. 

The first step was to define a beam-related background free ROI. While CCM is able to use the 10 $\mu$s of prebeam data collected to make highly accurate background rate predictions for the steady-state backgrounds such as cosmic rays, radioactivity, long lived neutron activation, etc, such prediction accuracy is not possible for the beam-related fast neutron backgrounds. As such a region free of such backgrounds is desirable. This region is defined as the time region starting with the earliest arrival of speed of light particles from the target created by the incident proton beam T$_0$ and ending with the so-called neutron beam-turn on. T$_0$ was determined through the use of direct measurement of speed of light particles from the gamma flash generated in the target with an EJ-301 scintillation detector on a dedicated beam line directly viewing the target. This procedure is described in more detail in ref.~\cite{CCM:2021leg} and \cite{thesis_Edward}. The end time was determined from changes in the efficiency of the selection cuts and the slope of the rate of high energy signal-like events, as shown in Fig.~\ref{fig:roiEnd}. The beam related neutron free ROI was determined from the data to be 150\,ns.  This is consistent with MCNP simulations for the earliest fast neutrons (kinetic energy $>100$\,MeV) arriving through the bulk shielding (5\,m steel and 2\,m concrete) and neutron time of flight measurements with multiple EJ-301 detectors in the vicinity of CCM behind the bulk shielding. This contribution is determined to be negligible.

The second step was the definition of the event selection cuts. A number of data quality cuts were used in all CCM120 searches, specifically a beam quality cut to ensure good quality beam triggers, a previous event cut to remove triplet events, and a veto cut to remove events from outside the detector. These cuts are described in more detail in ref.~\cite{CCM:2021leg}.

The signal selection cuts were added after these quality cuts and were defined based on calibration data using a $^{22}$Na $\gamma$ source (Fig.~\ref{fig:40nsCalibration}) and detector simulations (See Appendix~\ref{app:smearing}) of various energy electromagnetic events in CCM. The first such cut was a length cut requiring events >38 ns. Using a dynamic length event builder which defined conditions for the end of the event, the event length worked as a crude particle identification (PID) method. This PID was able to distinguish electromagnetic (electron or photon) events from nuclear recoil (neutrons or coherent scattering) events using the triplet light production of the former. This cut was >99\% efficient on ALP signals of all energies, and 66\% efficient on prebeam background. 

The second cut was a minimum energy cut from the high energy (>1 MeV) limits of the ALP search. The cut required events to have more than 10 PE of visible energy, approximately equivalent to 1 MeV at the center of the detector. This cut was $\sim95\%$ efficient on ALP signals above 1 MeV from simulations while being 30\% efficient on prebeam background, as the majority of the background was <1 MeV. The third cut was a strict position cut requiring radius <80 cm and height between -40 and 40 cm from the detector center (Fig.~\ref{fig:alpRadius}). This cut was $\sim75\%$ efficient on signal and 34\% efficient on background. 

Finally, two shape cuts were implemented requiring that the proportion of TPB coated PMTs seeing light was >0.6 and that the single PMT which saw the most light saw less than 20\% of the total charge (Fig~\ref{fig:maxPMT}). These two cuts combined were $\sim90\%$ efficient on signal and 40\% efficient on background. Over all reconstruction cuts, CCM120's selection criteria was $\sim65\%$ efficient on signal with variation according to the energy distribution of the signal (see Table 1), and flat 3.2\% efficient on background. 

\begin{figure}[h]
    \centering
    \includegraphics[width=\linewidth]{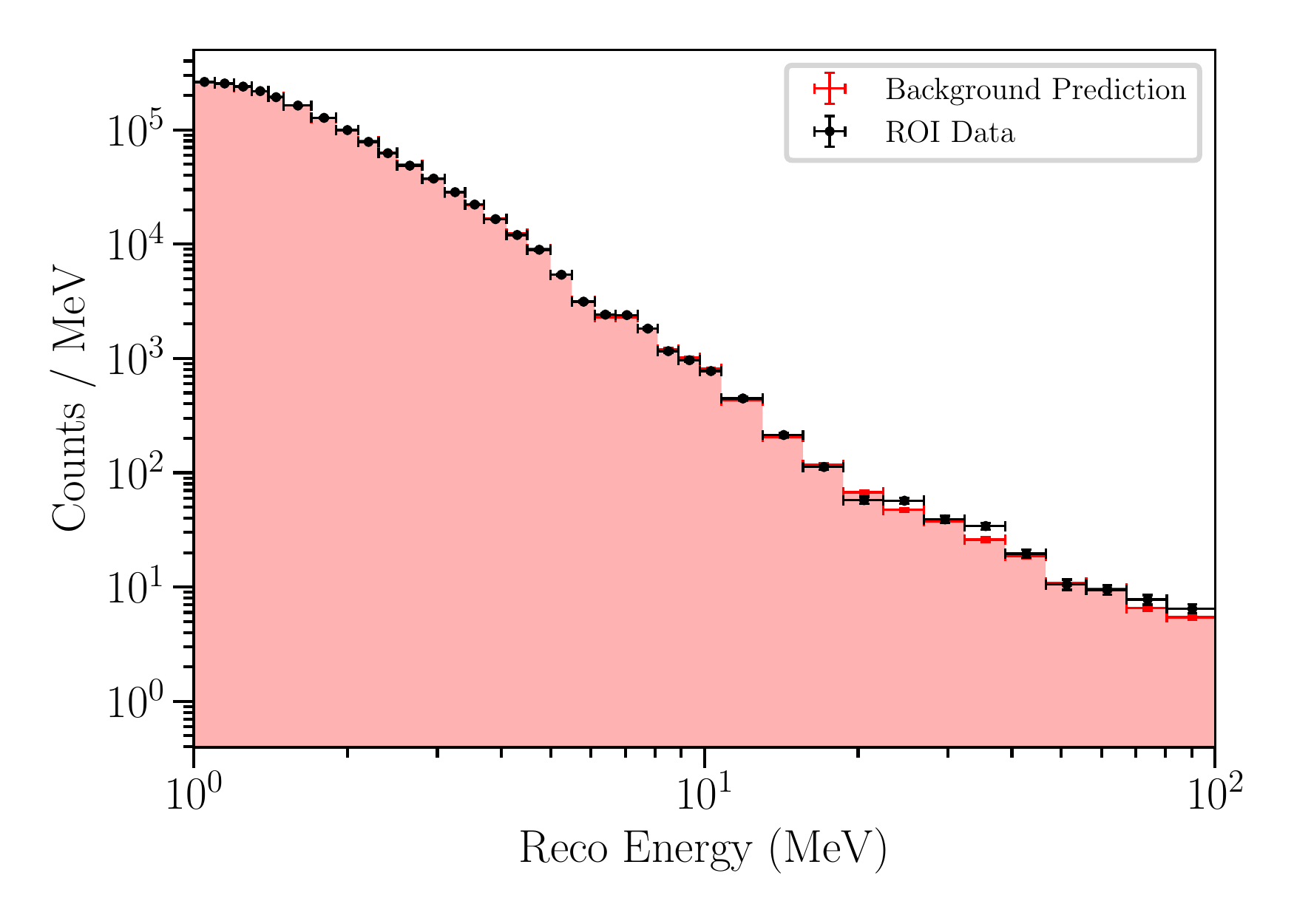}
    \caption{The CCM120 data and background spectra from the prebeam steady state background prediction and the measured data in the beam ROI, for $1.79\times10^{21}$ POT.}
     \label{fig:backgroundData}
\end{figure}

\begin{figure}[h]
    \centering
    \includegraphics[width=0.49\textwidth]{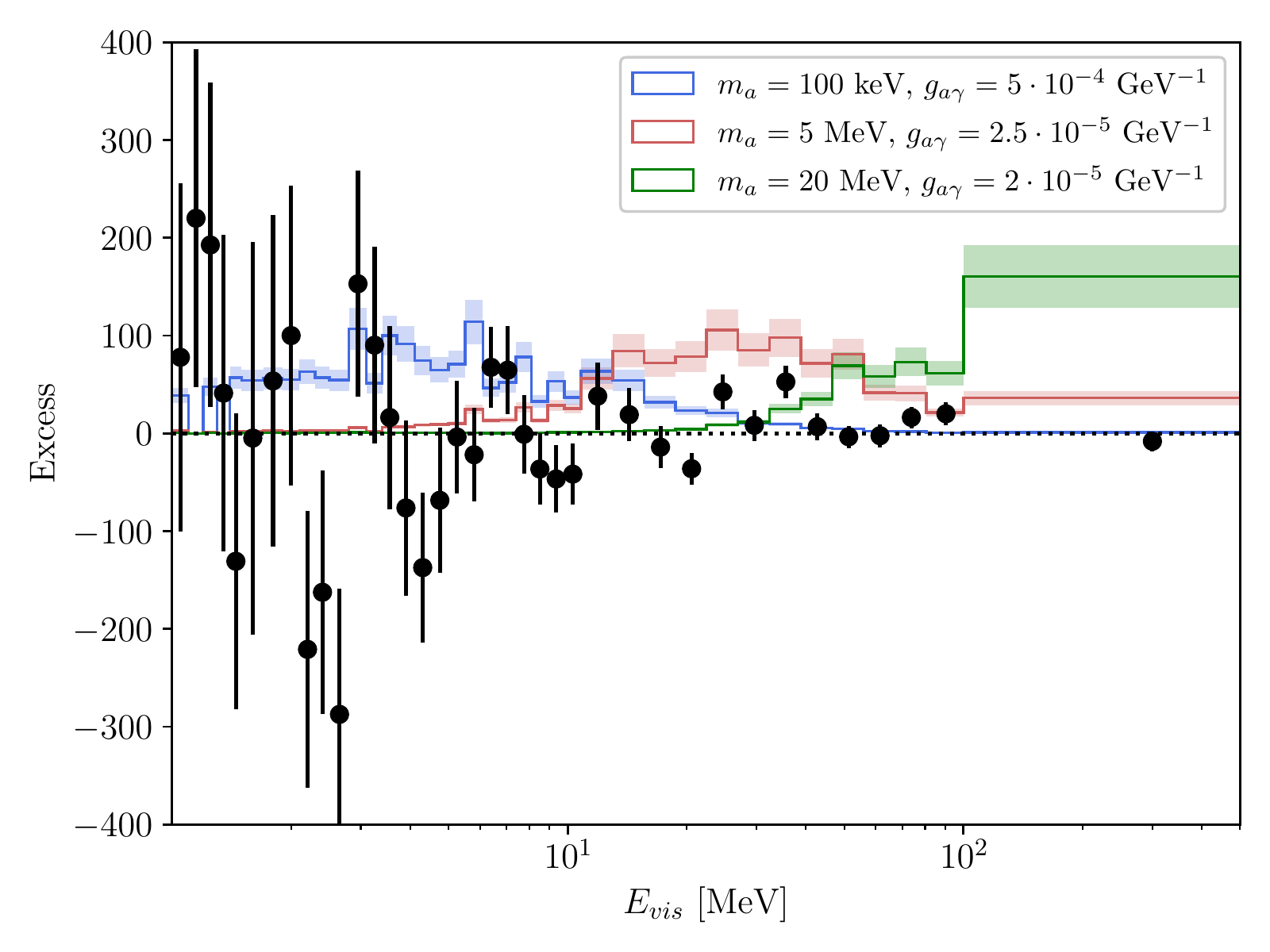}
    \includegraphics[width=0.49\textwidth]{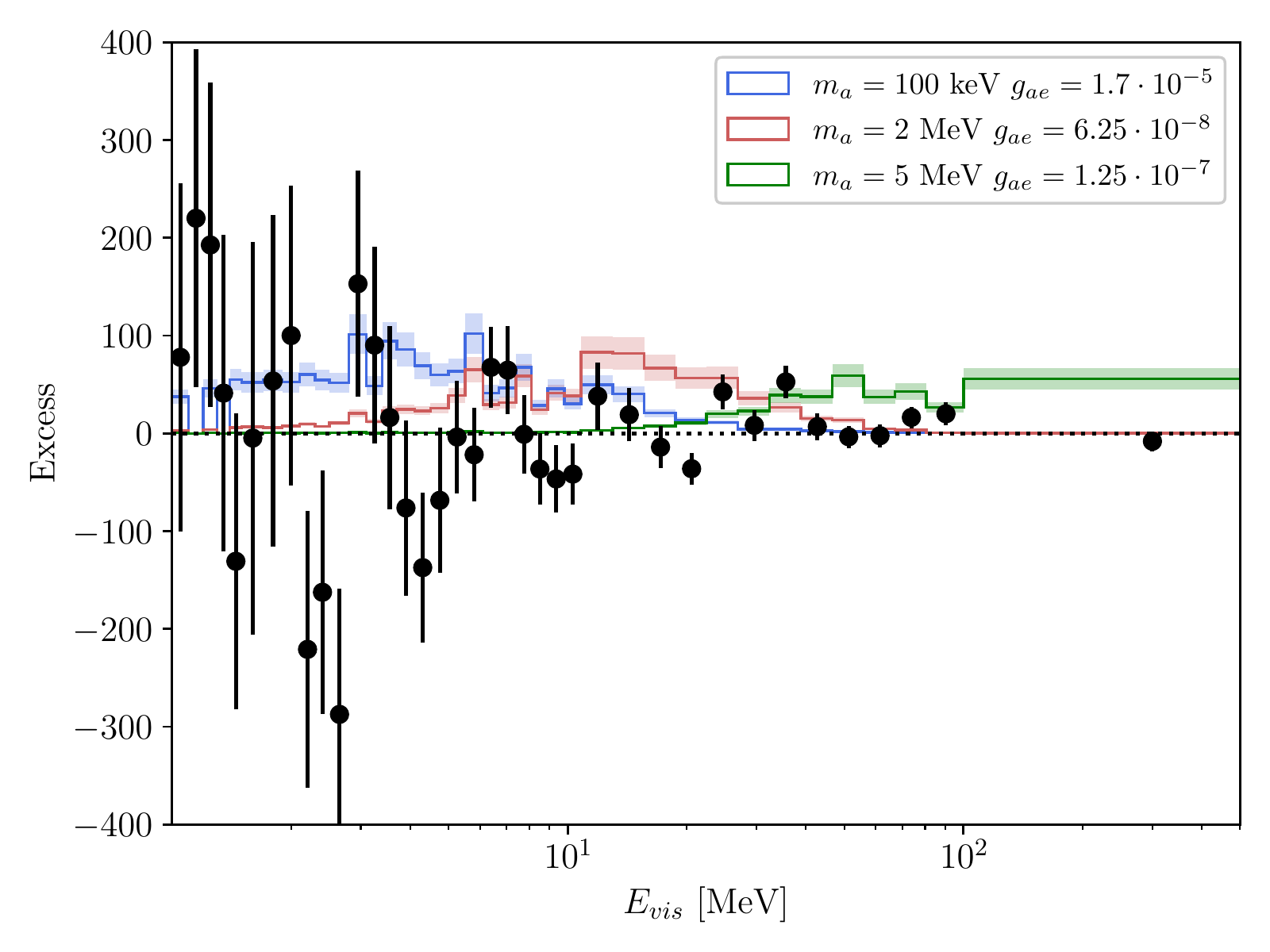}
    \caption{The subtraction reconstructed energy spectrum between data and background prediction for the CCM120 ALP search, compared to predicted event spectra at various masses for a photon (top) or electron (bottom) coupling (signal parameters chosen to show large event rates). Smearing uncertainties at $\pm 1\sigma$ are shown by the shaded bands. See Appendix~\ref{app:smearing} for details on the simulation and reconstruction.}
    \label{fig:subtract}
\end{figure}

The resulting data and background spectrum are shown in Fig.~\ref{fig:backgroundData}.  The number of data events in the signal region is $294590$, with a scaled prebeam steady state background prediction of $294614.3 \pm 241.7$ (syst) $\pm  542.8$ (stat) and a subtraction of $-24.3 \pm 594.2$. The error is statistics dominated. 
The systematic error on the steady state background prediction accounts for the additional variance from the systematic decay in background rates observed on the microsecond scale that are characteristic of radioactive decay from beam thermal neutron activation. After performing an exponential fit to account for this background, the error on the fit prediction is used as the systematic uncertainty on the background prediction. 
There is no significant excess over the entire energy range, however, axion signals can have strong energy dependence as shown in Fig.~\ref{fig:spectra} and require axion model fits to properly analyze. 
Since the $\chi^2$ of the data points in Fig.~\ref{fig:subtract} are consistent with no excess at the $2 \sigma$ level, there is no significant axion signal in the 2019 data, but the data can be used to determine exclusion regions as a function of model parameters. 


\begin{table*}[ht!]
\caption{ALP Search Efficiency Table.  The top two rows represent beam cuts, while the middle five rows represent detector reconstruction cuts.  The biggest improvement to efficiency would come from reducing the beam spill to less that 150\,ns, then making the signal selection time cut for speed of light particles 100\% efficient.}
\begin{center}
\label{efficiencyTable}
\begin{tabular}{l c c c c c} 
Cut & 1MeV ALP & 10MeV ALP & 20MeV ALP & 50 MeV ALP & Background \\
\hline
AllQuality & 0.749 & 0.886 & 0.936  & 0.969 & 0.149\\
Time & 0.393 & 0.447 & 0.451 & 0.458 & 0.0037\\
\hline
Length & 0.990 & 0.998 & 0.998 & 0.997 & 0.660\\
Energy & 0.933 & 0.990 & 0.991 & 0.992 & 0.202\\
Radius & 0.626 & 0.658 & 0.753 & 0.918 & 0.068\\
Coated & 0.616 & 0.656 & 0.751 & 0.917 & 0.066\\
maxPMT & 0.451 & 0.588 & 0.711 & 0.892 & 0.032\\ 
\hline
Total & 0.190 & 0.263 & 0.321 & 0.409 & 0.00012\\ [1ex] 
\end{tabular}
\end{center}
\end{table*}

\begin{figure}[h]
    \centering
    \includegraphics[width=0.49\textwidth]{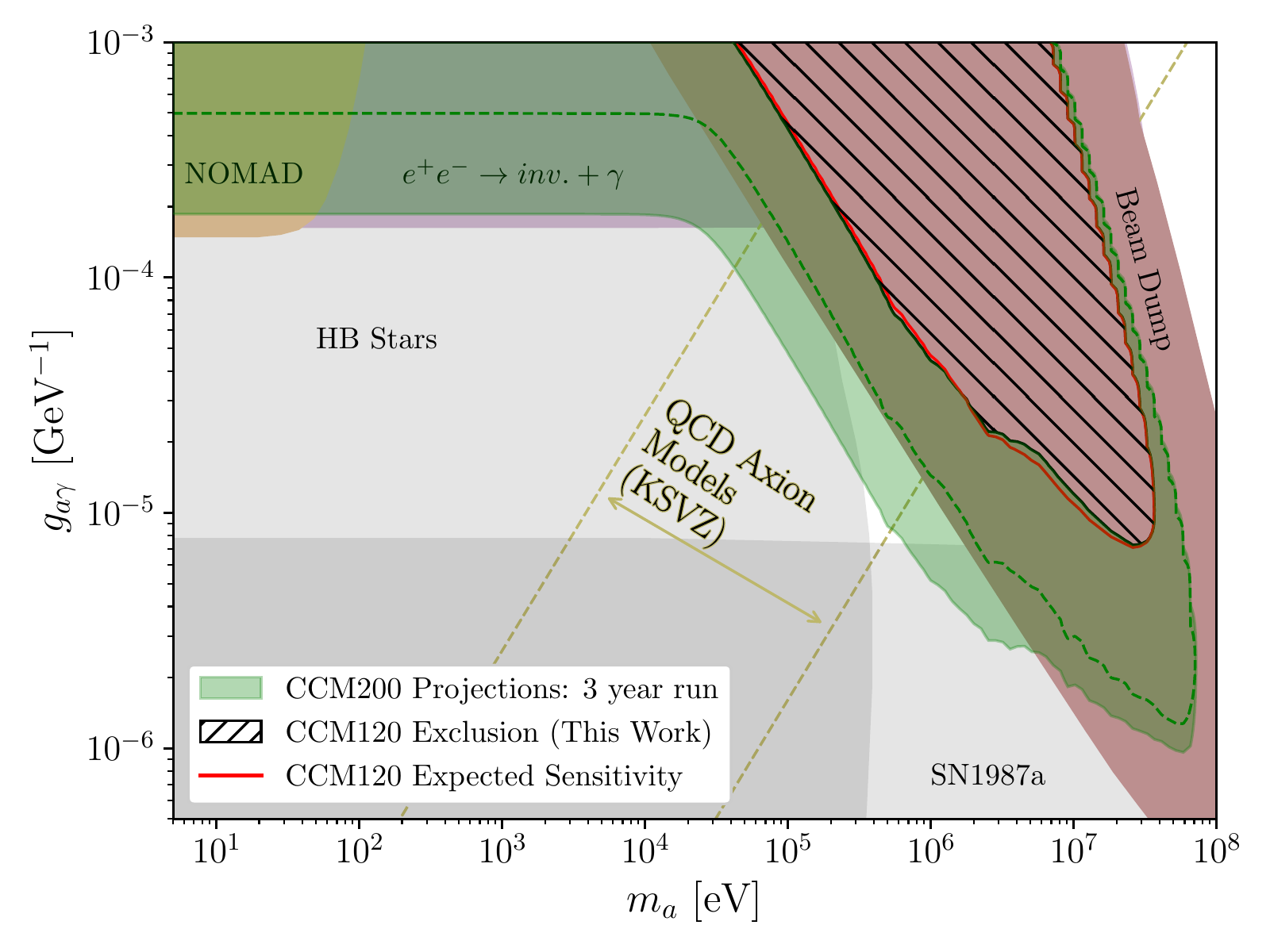}
    \caption{The expected and actual 90\% CLs from CCM120 for the ALP-photon coupling $g_{a\gamma}$. Also included is the projection region for CCM200 three year run using background taken from CCM120's spectrum reduced by two orders of magnitude for various conservative improvements (dashed green line) and a background free assumption (extent of shaded green region). QCD axion model parameter space for the KSVZ benchmark scenario spans the region indicated by the arrows~\cite{DiLuzio:2020wdo}.}
    \label{fig:limits120_photon}
\end{figure}

\begin{figure}[h]
    \centering
    \includegraphics[width=0.49\textwidth]{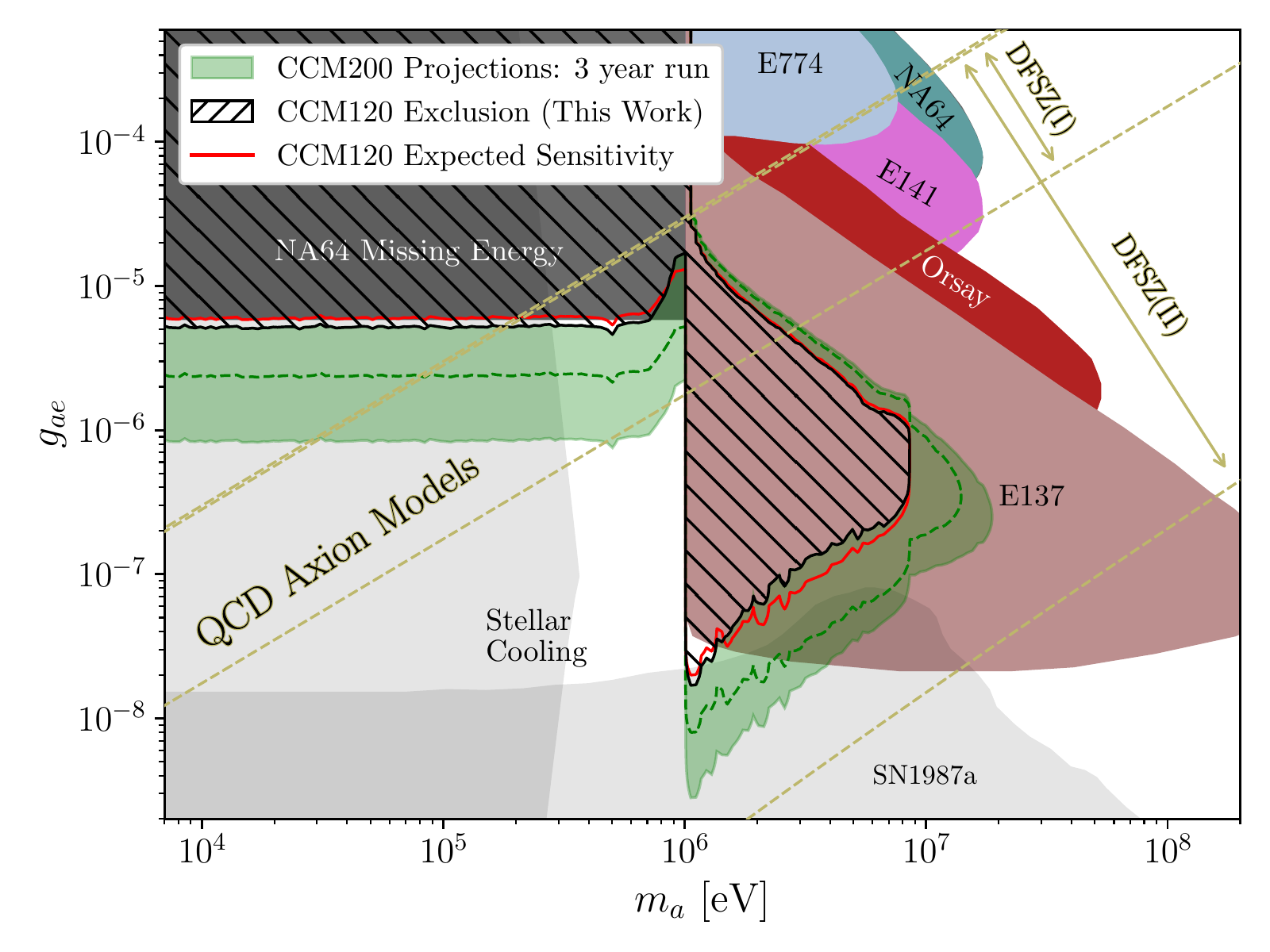}
    \includegraphics[width=0.49\textwidth]{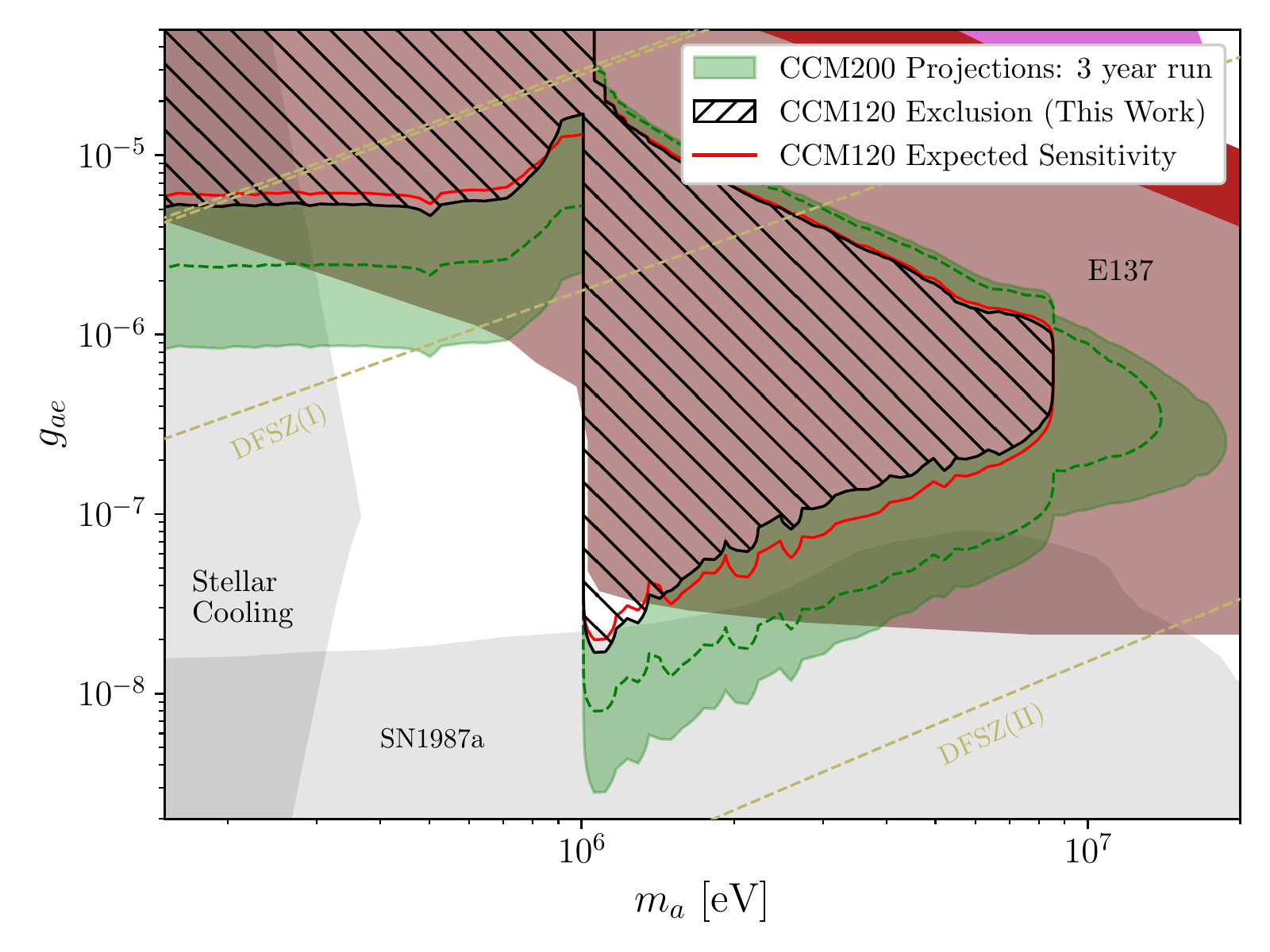}
    \caption{The expected and actual 90\% CLs from CCM120 for the ALP-electron coupling $g_{ae}$ at tree-level (top) and with an effective photon coupling at loop level (bottom). Also included are projections for CCM200, using the same background specifications as in Fig.~\ref{fig:limits120_photon}, for a 3-year run. QCD axion model parameter spaces for the DFSZ(I) and DFSZ(II) benchmark scenarios span the regions indicated by the arrows~\cite{DiLuzio:2020wdo}. The region excluded by missing energy searches at NA64 is shown in gray, and the bound derived this work from the CCM120 engineering run is set at marginally lower couplings than the NA64 region. Even in the conservative assumption that loop-level $a\to \gamma\gamma$ decays are not suppressed (bottom panel), CCM200 is projected to reach beyond the more stringent constraints set from E137 in this scenario.}
    \label{fig:limits120_electron}
\end{figure}

The third step in the CCM120 search was to use the data so selected to define a set of confidence limits. We defined these confidence limits as exclusion regions using a $\Delta \chi^2$ analysis and the asymptotic approximation given by Wilks' theorem. Using the ALP induced event rates (Fig.~\ref{fig:spectra}) for each ALP model, including mass and coupling, we defined a $\chi^2$ value for each point on the phase space map in the two ALP models, where the $\chi^2$ compared background ($\mathrm{bkgnd}_i$) and ALP prediction ($\mathrm{pred}_i$) to the data ($\mathrm{data}_i$) with respect to the total error $\sigma_i$ (stat+syst) across all reconstructed energy bins $i$ as seen in Eq.~(\ref{eq:confidencechi2}).

\begin{equation}
\label{eq:confidencechi2}
    \chi^2 = \sum_i \frac{(\mathrm{data}_i - \mathrm{bkgnd}_i - \mathrm{pred}_i)^2}{\sigma_i^2}
\end{equation}

This $\chi^2$ was used to evaluate the goodness of fit of the models shown in Fig.~\ref{fig:subtract}. We then defined $\Delta \chi^2_{model} = \chi^2_{model} - \chi^2_{best-fit}$ and from Wilk's theorem used the resultant $\Delta \chi^2$ values as a test statistic for a $\chi^2$ test with 2 degrees of freedom across the entire ALP parameter space. We could thus exclude phase space with $\Delta\chi^2>4.61$ to a 90\% confidence limit. We also calculated the experimental sensitivity using only background predictions, with an otherwise similar $\Delta \chi^2$ search methodology. The confidence limits and sensitivities so generated are shown in Figs.~\ref{fig:limits120_photon} and~\ref{fig:limits120_electron}.

For $g_{a\gamma}$ coupling, the existing beam-dump limits are better than the CCM120 exclusion limit as shown in Fig.~\ref{fig:limits120_photon}. For $g_{ae}$ coupling (as shown in Fig.\ref{fig:limits120_electron}), however, the CCM120 data constrains regions of parameter space in the $m_a \lesssim$ MeV mass range, which so far has only been probed by astrophysical observations. In some regions of the parameter space, the CCM constraint is found to be marginally better than both astrophysical and terrestrial constraints.

\section{Reach projections in ALP parameter space}\label{sec:reach}
The new and upgraded CCM200 detector (with 200 inward pointing PMTs, improved TPB foils, double veto PMTs, and LAr filtration (see ref.~\cite{CCM:2021leg}) was completed in 2021 and has begun a 3-year physics run from 2023 to 2025, which expects to collect an integrated luminosity of $2.25 \times 10^{22}$ POT. In this section we present projections for CCM200's sensitivity to the two ALP models discussed. These projections were obtained under two background scenarios: (i) a conservative one assuming background levels shown in Fig.\,\ref{fig:backgroundData} with an extra flat factor of 100 suppression for various improvements, and (ii) the background-free hypothesis.  The measured beam-on background distribution at CCM120 as a function of visible energy has been studied in \cite{CCM:2021leg}, and shown in Fig.~\ref{fig:backgroundData} for the ALP cuts (from ref.~\cite{thesis_Edward}). We conservatively assume that, with the CCM200 improvements in shielding, particle identification, and energy and spatial resolution, an additional factor of $\sim$100 in background rejection is possible. To first order this reduction is assumed to be flat in energy and has been demonstrated by data studies and simulations.  The ultimate goal is to go beyond this conservative estimate towards zero background with further improvements in added shielding, improved energy resolution with LAr filtration, reduced beam width by a factor of two or more, and analyses improvements in background rejection.

The 90\% C.\,L. projection reach of CCM200 in the parameter space of ALP coupling vs mass was obtained through simulations of pseudo-experiments under the background-only hypothesis and a standard Pearson $\Delta\chi^2$ test statistic.

Fig.~\ref{fig:limits120_photon} shows CCM200 sensitivity to an ALP model with electromagnetic couplings in the parameter space of ALP  $g_{a\gamma}$ coupling vs mass (see Eq.~(\ref{ALPlagrangian})). In the ALP mass range $m_a\lesssim 10$\,keV, the constraints from NOMAD~\cite{Astier:2000gx} and $e^+ e^- \to \gamma + \textrm{invisible states}$~\cite{Dolan:2017osp} are relevant. Here, the detection signal is dominated by inverse Primakoff scattering, and therefore CCM's reach is approximately mass-independent. At higher masses, where the signal is dominated by diphoton ALP decays, CCM200 will offer an improvement in reach over previous beam dump constraints, which is compelling since it will cover an unexplored region of ALP masses and couplings overlapping the KSVZ QCD axion parameter space~\cite{Kim:1979if,Shifman:1979if,DiLuzio:2020wdo}, including the so-called ``cosmological triangle'' ~\cite{Carenza:2020zil,Depta:2020wmr} formed by the boundary between astrophysical constraints~\cite{Jaeckel:2006xm,Khoury:2003aq,Masso:2005ym,Masso:2006gc,Dupays:2006dp, Mohapatra:2006pv,Brax:2007ak,DeRocco:2020xdt}, shown in gray, and the bounds from beam dumps~\cite{Jaeckel:2015jla,doi:10.1142/S0217751X9200171X}. 

Fig.~\ref{fig:limits120_electron} (top panel) shows CCM's reach to an ALP model with dominant electronic couplings in the parameter space of ALP $g_{ae}$ coupling vs mass (see Eq.~(\ref{ALPlagrangian})). Here, too, CCM200 will provide the strongest terrestrial constraints in the ALP mass range $m_a\leq2\,m_e$,
and will be able to probe parts of the QCD axion parameter space of DFSZ-I models\footnote{We point out, however, that the DFSZ axion is robustly excluded in the mass range of $\mathcal{O}(0.1-1)\;\text{MeV}$ due to upper-bounds on axion emission from decays of light mesons, quarkonia, and excited nuclear states \cite{ParticleDataGroup:2020ssz}.} \cite{osti_7063072,Dine:1981rt,1983PhLB..120..137D}. For masses within the range of $2\,m_e\lesssim m_a\lesssim 10\;\text{MeV}$, CCM will be able to extend the reach to smaller $g_{ae}$ couplings beyond the parameter space excluded by E137 \cite{PhysRevD.38.3375,Andreas:2010ms}, including a QCD axion corner belonging to DFSZ-II models. In this region, the only competing constraints come from Supernova SN1987a bounds ~\cite{Lucente:2021hbp}. Other constraints shown in Fig.\,\ref{fig:limits120_electron} were taken from \cite{Bechis:1979kp} (Orsay), \cite{Riordan:1987aw} (E141), \cite{Bross:1989mp} (E774), \cite{NA64:2021ked,Andreev:2021fzd,Gninenko:2017yus} (NA64), and \cite{Hardy:2016kme} (stellar cooling).

In Fig.~\ref{fig:limits120_electron} (bottom panel) we show the case in which an effective ALP-photon coupling is generated through an electron loop, permitting $a \to \gamma\gamma$ decays. This modifies the parameter space such that E137 constrains lower mass parameter space below $m_a < 2 m_e$ from sensitivity to the $\gamma\gamma$ decay channel~\cite{PhysRevD.96.016004}. In this more conservative scenario, where loop-level $a\to \gamma\gamma$ decays are not suppressed, CCM200 is projected to reach beyond the more stringent constraints set from E137 in this scenario.

\section{Conclusion}\label{sec:conclusion}
This work illustrates the breadth of physics signatures that the CCM experiment will be able to explore. Being a proton beam experiment, CCM is uniquely suited to probe new physics signals initiated from hadronic processes. However, the high intensity of photons and electromagnetic cascades generated at the source will also allow CCM to probe new light particles that are coupled electromagnetically or electronically. A specially well-motivated class of models in this category are axion-like particles, which encompass the QCD axion that solves the strong CP problem, but also more generic light pseudoscalars from a dark sector.

In particular, given its unprecedented sensitivity to a lower energy kinematic range, CCM will be able to probe a variety of softer dark sector signatures ranging from coherent elastic dark matter/neutrino nuclear scattering, to $\mathcal{O}(0.1-100)\;\text{MeV}$ electromagnetic signals. This enables CCM to test a range of lighter and more weakly-coupled ALPs that was outside the reach of previous generations of beam dump/fixed target experiments. For the specific ALP models considered in this study, CCM can provide the leading terrestrial constraints in the sub-MeV ALP mass range. In this range, competing constraints stem mostly from bounds on stellar cooling, which are more model dependent and subject to large astrophysical uncertainties. Intriguingly, CCM can also probe surviving corners of the QCD axion parameter space in the $m_a\sim \mathcal{O}(0.1-1)\;\text{MeV}$ range that have proven difficult to exclude with past experiments and astrophysical observations.

In this paper, we analyzed the 2019 engineering run data for both $g_{a\gamma}$ and $g_{ae}$ couplings. We found  that the CCM data has constrained regions of parameter space in the $\lesssim$ MeV mass range for the $g_{ae}$ coupling  which so far has only been probed by astrophysical 
observations and  in some regions of the parameter space, the CCM constraint is found to be marginally better than  both astrophysical and terrestrial constraints. We also showed the projected sensitivity for a three year run and utilizing a  background reduction of greater than two orders-of-magnitude relative to the baseline measured during CCM's 2019 engineering run. With CCM200 ongoing
detector performance improvements we expect to achieve these background levels starting in 2022 and with a 3-year physics run from 2023 to 2025 (collecting $2.25 \times 10^{22}$ POT) will achieve the sensitivities stated in this work. In the longer term, future planned upgrades that shorten the beam pulse width by an order of magnitude could provide even further background rejection and increased signal sensitivity. In the event of a discovery or observation of an excess, CCM's ability to detect visible energy (instead of missing energy/momentum) would also offer an advantage in narrowing down the possible BSM explanations for a putative signal.

\section*{Acknowledgements}

 We acknowledge support from the Department of Energy Office of Science, the National Science Foundation, Los Alamos National Laboratory LDRD program, the National Laboratories Office at Texas A\&M University, and PAPIIT-UNAM grant No. IT100420. Portions of this research were conducted with the advanced computing resources provided by Texas A\&M High Performance Research Computing. We also wish to acknowledge support from the LANSCE Lujan Center and LANL's Accelerator Operations and Technology (AOT) division.  This research used resources provided by the Los Alamos National Laboratory Institutional Computing Program, which is supported by the U.S. Department of Energy National Nuclear Security Administration under Contract No.\,89233218CNA000001.

\bibliography{main}
\bibliographystyle{unsrt}

\onecolumngrid
\appendix

\section{Fluxes and event rates for ALP production and detection}
\label{app:cross_sections}

In this appendix, we review the scattering and decay amplitudes relevant for axion-like particle production and detection. Most of the cross sections and decay widths are already given by various references in the literature, for which we simply collect them here for the convenience of the reader; otherwise, we explicitly provide matrix element and phase space integration information for channels we have computed ourselves.

\subsection{ALP-electron coupling: Production and Detection Mechanisms}
ALPs coupling to electrons via the pseudoscalar Yukawa operator,
\begin{equation}
    \mathcal{L} \supset -i g_{ae} a \bar{e} \gamma^5 e ,
\end{equation}
have many production and detection channels available to them that resemble similar scattering processes of the SM photon. The production and detection channels that are relevant for stopped-pion beam target experiments with this coupling are itemized below:
\flushleft{$\diamond$ \textit{Compton scattering}:}\\
The Compton-like scattering process for ALP production from photons scattering on electrons at rest, $\gamma + e^- \to a + e^-$, is shown in Fig.~\ref{fig:axionCompton}. It has a differential scattering cross-section, in light-cone coordinates,
\begin{equation}
    \frac{d\sigma_C}{dx} = \frac{Z \pi g_{ae}^2 \alpha x}{4\pi(s- m_e^2)(1-x)}\bigg[x - \frac{2m_a^2 }{(s-m_e^2)^2} \bigg(s - \frac{m_e^2}{1-x} -\frac{m_a^2}{x}\bigg)\bigg]
    \label{eq:compton}
\end{equation}
where $x = 1 - E_a / E_\gamma + m_a^2 / (2E_\gamma m_e)$ is the energy fraction distributed to the ALP in light-cone coordinates. This formula was presented in ref.~\cite{Brodsky:1986mi}, although we have corrected a sign mistake for the second term in square brackets in Eq.~\ref{eq:compton}, also pointed out in ref.~\cite{AristizabalSierra:2020rom}. The differential cross section with respect to the outgoing ALP energy can be obtained by taking $d\sigma/dE_a = (1/E_\gamma) d\sigma/dx$, and the relevant bounds of integration over $x$ can be taken from ref.~\cite{AristizabalSierra:2020rom} and references therein. 

\flushleft{$\diamond$ \textit{Inverse Compton Scattering}:}\\
The inverse-Compton scattering process ($a + e^- \to \gamma + e^-$) shown for ALP detection, shown in Fig.~\ref{fig:axionInvCompton}, produces visible energy in the final state electron recoil and outgoing photon. The total cross section is~\cite{Avignone:1988bv,Gondolo:2008dd};
\begin{equation}
    \sigma(E_a) = \dfrac{g_{ae}^2 \alpha}{8 m_e p_a} \bigg[ \frac{2 m_e^2 (E_a + m_e) y}{(y + m_e^2)^2} + \frac{4 m_e (m_a^4 + 2m_e^2 m_a^2 - 4m_e^2 E_a^2)}{y (y + m_e^2)} + \frac{4 m_e^2 p_a^2 + m_a^4}{p_a y} \ln \bigg(\frac{m_e + E_a + p_a}{m_e + E_a - p_a} \bigg) \bigg]
\end{equation}
where $y = m_a^2 + 2 m_e E_a$.

\flushleft{$\diamond$ \textit{Associated Production}:}\\
The associated production of an ALP with a $\gamma$ via electron-positron annihilation is shown in Figure~\ref{fig:axionAssociated}, with momenta $e^+ (p_+) e^-(p_-) \to a(k) + \gamma(q)$. The matrix element is computed below;
\begin{equation}
    \mathcal{M} = -i g_{ae} e \bigg(\frac{[\bar{v}(p_+)\gamma^5 (\slashed{p}_+ - \slashed{k} + m_e)\gamma^\mu \epsilon^*_\mu (q) u(p_-)]}{t-m_e^2} + \frac{[\bar{v}(p_+)\gamma^5 (\slashed{p}_+ - \slashed{q} + m_e)\gamma^\mu \epsilon^*_\mu (q) u(p_-)]}{u-m_e^2}\bigg)
\end{equation}
\begin{align}
    \braket{|\mathcal{M}|^2} &= 4\pi \alpha g_{ae}^2 \bigg[\frac{3 m^4-m^2 (m_a^2+s)+t (-m_a^2+s+t)}{(m^2-t)^2} \nonumber \\
    &+ \frac{7 m^4+m^2 (m_a^2-3 s-4 t)+t (-m_a^2+s+t)}{(m^2+m_a^2-s-t)^2} \nonumber \\
    &+2\frac{3 m^4-m^2 (s-2 t)+t (m_a^2-s-t)}{(m^2-t) (m^2+m_a^2-s-t)} \bigg]
\end{align}
The differential cross section is then given by the general expression for $2\to 2$ scattering,
\begin{equation}
        \dfrac{d\sigma}{dt} = \dfrac{1}{16\pi(s-(m_1 + m_2)^2)(s-(m_1 - m_2)^2)} \braket{|\mathcal{M}(s,t)|^2}  ,
\end{equation}
which gives, in the CM frame (denoted with starred variables)
\begin{equation}
\label{eq:alp_mc}
    \dfrac{d\sigma}{d(\cos\theta^*)} = 2 p^*_1 p^*_3 \dfrac{d\sigma}{dt}
\end{equation}
Weighted samples of ALP 4-vectors can be drawn from this distribution, for example, by randomly drawing angles on the 2-sphere for the ALP in the CM frame, then transforming the resulting 4-vectors to the lab frame via a Lorentz boost. The MC weights are then given by $d\sigma$ which is frame invariant in the limit of large sample size.

\flushleft{$\diamond$ \textit{Axion Bremsstrahlung Production}:}\\
ALP bremsstrahlung ($e^\pm + N \to e^\pm + N + a$) shown in Fig.~\ref{fig:axionBrem} via the interaction of electrons or positrons with the strong nuclear electric field of atoms in material was studied by Tsai~\cite{PhysRevD.34.1326}. The differential cross section as a function of outgoing ALP energy $E_a$ in the Weizsacker-Williams approximation is given as
\begin{align}
    \dfrac{d\sigma}{dE_a} &= \frac{r_0^2 g_{ae}^2}{2 \pi E_+} \bigg[ \frac{x (1 + \frac{2}{3}f)}{(1+f)^2}\big( Z^2 \ln(184 Z^{-1/3}) + Z \ln (1194 Z^{-2/3}) \big) \nonumber \\
    &+ x \bigg( \frac{1}{3 f^2} (1 + f) \ln (1+f) - \frac{1 + 4f + 2f^2}{3 f(1+f)^2} \bigg)(Z^2 + Z) \bigg]
\end{align}
for $f = (m_a/m_e)^2 (1- x)/x^2$, $r_0 \equiv \alpha / m_e$.

\flushleft{$\diamond$ \textit{Resonant Production}:}\\
The resonantly produced ALP from positrons annihilating on electrons at rest ($e^+ e^- \to a$, Fig.~\ref{fig:axionProductionResonance}) has a final state energy of
\begin{align}
    E_a = \dfrac{m_a^2}{2m_e}.
\end{align}
This by itself is an interesting result, as it implies that the induced $a$ flux from electron-positron annihilation will peak strongly close to $\sim m_a^2$ (MeV).

The cross section in the electron rest frame with the narrow width approximation is
\begin{align}
    \sigma &= \dfrac{2 \pi m_e g_{ae}^2 s}{m_a^2 \sqrt{s(s-4m_e^2)}} \delta(E_+ - (\frac{m_a^2}{2m_e} - m_e))
\end{align}
where $s = m_e^2 + 2E_+ m_e$.

\flushleft{$\diamond$ \textit{ALP external pair production}:}\\
The process $a(k) N(p_1) \to N(p_2) e^+(\ell_+) e^-(\ell_-)$ shown in Fig.~\ref{fig:axionPairProduction}, analogous to photon-lepton pair production in the SM, was computed in refs.~\cite{KIM198387, KIM1984189} using the formalism and atomic form factors presented in ref.~\cite{RevModPhys.46.815}. The cross section may be expressed in terms of invariants $s = (k + p_1)^2$, $s_{\ell\ell} = (\ell_+ + \ell_-)^2$, and $t = (p_1 - p_2)^2$ and integrated in the dilepton CM frame;
\begin{equation}
    \dfrac{d^4 \sigma}{dt d s_{ll} d\Omega_{ll}^{CM}} = \frac{\alpha^2 g_{ae}^2 \beta}{64\pi^2 (2 M E_a)^2 t^2} L_{\mu\nu} H^{\mu\nu}
\end{equation}
Here we define the leptonic and hadronic tensors;
\begin{align}
    L_{\mu\nu} &= \text{Tr} \bigg[ \left(\slashed{\ell}_-+m\right) \left(\gamma_\mu \frac{\slashed{k}-\slashed{\ell}_+ + m}{m_a^2-2 \left(k\cdot \ell_+\right)} \gamma^5+\gamma^5 \frac{\slashed{\ell}_- - \slashed{k} + m}{m_a^2-2 \left(k\cdot \ell_- \right)} \gamma_\mu\right) \\
    & \times \left(\slashed{\ell}_+ - m\right)
    \left(\gamma^5 \frac{\slashed{k}-\slashed{\ell}_+ +m}{m_a^2-2 \left(k\cdot \ell_+ \right)} \gamma_\nu + \gamma_\nu \frac{-\slashed{k}+\slashed{\ell}_- +m}{m_a^2-2 \left(k \cdot \ell_- \right)} \gamma^5\right) \bigg]
\end{align}
\begin{equation}
H^{\mu\nu} = \frac{1}{2} \text{Tr} \bigg[\left(\slashed{p}_2 + M\right).\gamma^\mu \left(\slashed{p}_1 + M\right).\gamma^\nu \bigg] \times F_A^2 (t)
\end{equation}
where $F_A(t)$ is the atomic form factor described in ref.~\cite{RevModPhys.46.815}.

\flushleft{$\diamond$ \textit{Decays to electron-positron pairs}:}\\
Decays to $e^+ e^-$ pairs are permitted for ALP masses $m_a > 2 m_e$, i.e., $m_a \gtrsim 1$ MeV. This decay is pictured in Fig.~\ref{fig:axionDecayElectronPositron}, and the decay width is given by Eq.~\ref{eq:electron_width};
\begin{equation}
    \Gamma(a \to e^+e^-) = \frac{g_{ae}^2 m_a}{8\pi}\sqrt{1 - \frac{4m_e^2}{m_a^2}}.
    \label{eq:electron_width}
\end{equation}

\subsection{ALP-photon coupling Production and Detection Mechanisms}
ALPs coupling to photons via the dimension-5 operator
\begin{equation}
    \mathcal{L} \supset \frac{g_{a\gamma}}{4} a F_{\mu\nu}\Tilde{F}^{\mu\nu},
\end{equation}
give rise to production and detection channels from Primakoff scattering as well as decays to two photons. The amplitudes for these processes are described below.

\flushleft{$\diamond$ \textit{Primakoff and Inverse Primakoff Scattering}:}\\
The matrix element for Primakoff scattering with a free, heavy fermion $a(k) N(p) \to \gamma(k^\prime) N(p^\prime)$ with momentum transfer $q = k - k^\prime$ is depicted in Fig.~\ref{fig:axionPrimakoff}. The matrix element for scattering off a free fermion plane wave state is
\begin{equation}
    \mathcal{M}_\textrm{free} = \Bar{u}(p^\prime) (-i e \gamma^\mu) u(p) \bigg(\dfrac{-ig_{\mu\nu}}{q^2} \bigg) (i g_{a\gamma} \epsilon^{\nu\rho\alpha\beta} q_\alpha k^\prime_\beta) \varepsilon^*_\rho (k^\prime)
\end{equation}
Squaring $\mathcal{M}_\textrm{free}$ and evaluating the trace in terms of the usual mandelstam variables $t = -q^2$ and $s = (k + p)^2$ yields
\begin{equation}
    \braket{\mid \mathcal{M}\mid^2}_\textrm{free} = \frac{g_{a\gamma}^2 t (2 M^2 (m_a^2-2 s-t)+2 M^4-2 m_a^2 (s+t)+m_a^4+2 s^2+2 s t+t^2)}{8 t^2}
    \label{eq:m2ps}
\end{equation}
To apply this to a real atomic target, we can factorize the free matrix element and the atomic form factor separately to compute the interaction with the nuclear and electron cloud charge density in a straightforward way, as
\begin{equation}
    \braket{\mid \mathcal{M}\mid^2} = \braket{\mid \mathcal{M}\mid^2}_\textrm{free} \times F^2(q)
\end{equation}
For a hydrogenic potential, we use Tsai's parameterization~\cite{PhysRevD.34.1326} of the atomic form factor $F^2(t)$,
\begin{equation}
    F_A^2(q) = \dfrac{Z^2 a^4 q^4}{(1+a^2 q^2)^2}
\end{equation}
with $a=184.15 e^{-1/2} Z^{-1/3} / m_e$.

For inverse Primakoff scattering, the cross section is twice as big (since we no longer average over initial state photon polarizations) but is otherwise the same, with the appropriate re-definitions of Mandelstam invariant $s = 2 E_a M + M^2 + m_a^2$.
If we instead use a simplified dipole form factor with screening length $r_0$ and express the differential cross section in the forward limit, the total cross section can be compactly expressed as~\cite{Avignone:1988bv,Creswick:1997pg,Avignone:1997th}
\begin{equation}
    \sigma(k_a) = \dfrac{Z^2 \alpha g_{a\gamma}^2}{2} \bigg(\dfrac{2 r_0^2 k_a^2 + 1}{4 r_0^2 k_a^2} \ln \Big(1+4r_0^2 k_a^2 \Big) - 1 \bigg),
    \label{eq:xs}
\end{equation}
which can be also used for $\gamma \to a$ Primakoff scattering in the massless ALP limit after dividing by a factor of 2 to account for the polarization averaging.

\flushleft{$\diamond$ \textit{Decays to photon pairs}:}\\
The decay process $a \to \gamma \gamma$ pictured in Fig.~\ref{fig:axionDecayDiphoton} is kinematically available at all ALP masses, although this decay is most relevant for masses in the MeV-GeV mass range at the energy and length scales of beam dump and stopped pion facilities. The width is
\begin{equation}
    \Gamma(a \to \gamma\gamma) = \dfrac{g_{a\gamma}^2 m_a^3}{64 \pi}
\end{equation}

\subsection{MCMC Method for Event Rate Distributions at CCM}
We use an internal python library containing the matrix elements and cross sections listed above for input into a MCMC event generator. A schematic breakdown of the simulation pipeline is as follows:
\begin{enumerate}
    \item Pass in fluxes of $e^\pm$, $\gamma$ transport inside the tungsten target from \texttt{GEANT4} simulation as arrays of energies (MeV) and weights (counts $\cdot$ POT$^{-1}$).
    \item For each $e^+$, $e^-$, $\gamma$ generate a weighted spectrum of ALPs using the appropriate production channel cross sections (for example, using the procedure outlined under Eq.~\ref{eq:alp_mc}).
    \item Assume the ALP flux produced is isotropic, propagating the flux to the detector with a suppression factor of $1/(4\pi \ell^2)$ for $\ell = 23$ m (CCM120).
    \item Generate MCMC sampled distributions from the ALP fluxes propagated to the detector, appropriately reweighted to get the expected counts per exposure and visible energies in the ROI.
    \item Convolve the distribution with the smearing matrix derived from the optical model and selection efficiencies described in Appendix~\ref{app:smearing}.
\end{enumerate}

First, we review general formulae for predicting isotropic ALP fluxes produced inside the tungsten target. 

Photons produced and transported through the tungsten target are assumed to be entirely absorbed, and with minimal energy loss before getting stopped. In this case, the ALP production rate can be modeled by taking a ratio of the SM photon absorption cross section $\sigma_\gamma$ and the ALP production cross section $\sigma_a$;
\begin{equation}
    \dfrac{dN_a}{dE_a dt} = \int  \dfrac{dN_\gamma}{dE_\gamma dt}\dfrac{1}{\sigma_\gamma (E_\gamma)}\dfrac{d\sigma_a(E_\gamma)}{dE_a} dE_\gamma
\end{equation}

For ALPs produced from resonant production, bremsstrahlung, and associated production, the energy loss of the electrons and positrons in material must also be folded in. In the case of resonant production, this modifies the ALP production rate as 
\begin{align}
    \dfrac{dN_a}{dt} = \frac{X_0 N_A}{A} (\hbar c)^2 \int_{E_+^{min}}^{E_+^{max}} \int_{E_+^\textrm{min}}^{E_+} \int_0^T\frac{dN_{e^+}}{dE_+dt} I(t, E_+, E^\prime) \sigma_a(E^\prime) dt dE^\prime dE_+
\end{align}
where the track length probability function
\begin{equation}
I(t, E_i, E_f) = \frac{\theta(E_i - E_f)}{E_i \Gamma (4 t/3)} (\ln E_i/E_f)^{4t/3 - 1}
\end{equation}
is the energy loss smearing function for the electron/positron radiation length $t$ and target radiation thickness $T$~\cite{PhysRevD.34.1326}. The prefactor accounts for the radiation length $X_0$ in g/cm$^2$, the atomic weight of tungsten $A$ in g/mol, and Avogadro's number $N_A = 6.023 \cdot 10^{23}$. For resonant production, the delta function in $\sigma(E^\prime)$ removes the $dE^\prime$ integral and fixes the final state energy to the resonant energy. In the case of associated or bremsstrahlung production, we replace $\sigma (E^\prime)$ with $d\sigma/dE^\prime$ and numerically carry out the integration over $dt dE^\prime dE_\pm$.

The propagation of these fluxes isotropically to a distance $\ell$ away from the production site will multiply the number event rate $dN_a/dt$ by a $1/(4\pi\ell^2)$ factor. In the limit that the distance $\ell$ from the source to the detector is much larger than the detector size, we can approximate the detector as a thin shell covering an area $\Omega_d$ with thickness $\Delta\ell$. In this limit we treat $\ell$ as a constant over the transverse directions $(\theta, \phi)$ over a sphere of length $\ell$ surrounding the source. The event rates for scattering and decays in terms of the ALP number rates produced in the W target, the detector particle number density $n$, and visible energy $E_{vis}$ can then be expressed simply as
\begin{align}
    \textrm{Scattering: } \dfrac{d^2 R}{dE_{vis} dt} &= \int \dfrac{d^2N_a}{dE_a dt} \bigg[ \frac{\Omega_d}{4\pi\ell^2}  P_{surv}(\ell) \frac{d\sigma(E_a)}{dE_{vis}} n  \ell \bigg] dE_a \label{eq:scatter_er}\\
    \textrm{Decays: }\dfrac{d^2 R}{dE_{vis} dt} &= \dfrac{d^2N_a}{dE_a dt} \frac{\Omega_d}{4\pi\ell^2} P_{decay}(\ell) \label{eq:decay_er}
\end{align}
The visible energy $E_{vis}$ is defined for scatterings (e.g. $a + e^- \to \gamma + e^-$) by the differential scattering cross section $d\sigma/dE_{vis}$, and we have assumed for $a\to \gamma \gamma$ and $a\to e^+ e^-$ decays that the incoming ALP energy is totally converted inside the detector volume (e.g. $E_a = E_{vis}$), before taking into account the smearing effects described in Appendix~\ref{app:smearing}. Here we have used the ALP lifetime in the laboratory,
\begin{equation}
    \tau_{lab} = \dfrac{1}{\Gamma_a} \times \dfrac{E_a}{m_a}
\end{equation}
and the exponential law for decays,
\begin{equation}
    p(z) = \frac{1}{\tau_{lab} v_a} e^{-z /\tau_{lab} v_a }   
\end{equation}
which together give us the probability that the ALP will survive until a distance $\ell$ from its production site,
\begin{equation}
    P_{surv}(\ell) = 1 - \int_0^\ell \frac{1}{\tau_{lab} v_a} e^{-z /\tau_{lab} v_a } dz = e^{-\ell / \tau_{lab} v_a}
\end{equation}
\begin{equation}
    P_{decay}(\ell) = e^{-\ell / \tau_{lab} v_a}(1- e^{-\Delta\ell / \tau_{lab} v_a})
\end{equation}
We then integrate the energy event rates given in Eq.~\ref{eq:scatter_er} and Eq.~\ref{eq:decay_er} over the energy bins ROI and time exposure, convolving the true energy $E_{vis}$ with the smearing matrix described in Appendix~\ref{app:smearing}, to get the predicted signal events at CCM. 

\section{Description of the Detector Optical Model, efficiencies, and the construction of the smearing matrix}\label{app:smearing}

The Optical Model (OM) was the portion of the simulation that defined the optical properties of the materials in which the photons would interact. 
These materials were the liquid argon (LAr), PMT glass, mylar foils, and Tetraphenyl Butadiene (TPB).
The OM was built primarily from 2 sets of calibration data: the $\gamma$-emitting $^{22}$Na and $^{57}$Co sources and a 2-frequency laser at 213 and 532 nm wavelengths. 
We used the two frequencies of the laser to probe the difference in detector response to UV and visible wavelengths of light. 
However, due to the significant difference in photon energy and thus detector response between the laser UV - 213 nm - and the argon scintillation peak at 128 nm, we also used the radioactive sources to probe the detector behavior at that scintillation peak. 

\begin{figure}[h]
    \centering
    \includegraphics[width=0.48\textwidth]{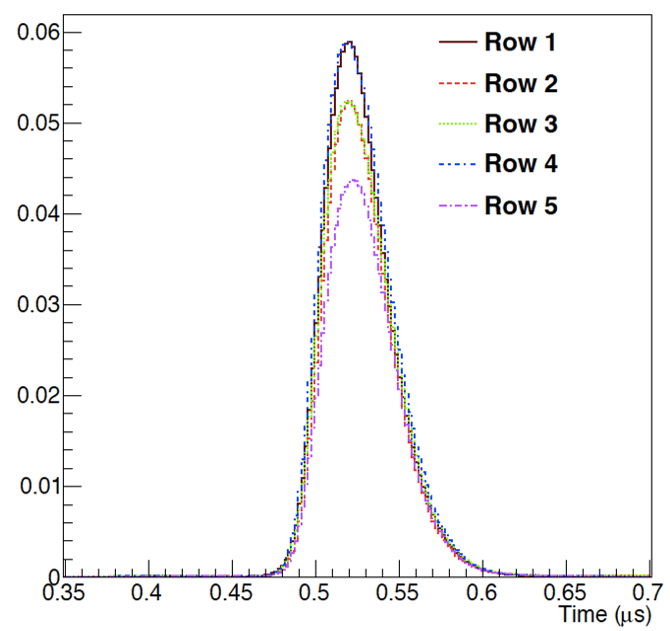}
    \caption{The integral charge from laser pulses in each of the PMT rows, used for calculating the row ratios used for determining the OM. The difference in internal position and distance from the source between the rows was used to calibrate distance dependent OM properties.}
    \label{fig:rowRatios}
\end{figure}

As the optical properties were primarily distance dependent, a method of obtaining this dependence was developed using the differing PMT positions of the height differentiated rows. 
Defining row 1 as that nearest the top of the detector, we created a set of 4 row ratios for each calibration run. 
These ratios were defined as the total light seen by each of the other 4 rows over that seen by row 1 over the course of a run. An example of the raw values used for this calculation are shown in Figure \ref{fig:rowRatios}. 
Between the two sources and five laser positions and two frequencies and with the addition of a fifth internal ratio - light seen by uncoated tubes versus that of coated - this gave us a total of 60 row ratios to use for the calibrations of the OM. 
These row ratios were used to compare between the data and simulation outputs through a $\chi^2$ analysis \cite{thesis_Edward}.

The process used for determining the parameter values involved generating simulations with varying OM parameters. 
Minimizing the $\chi^2$ over the row ratios of a simulation would be used to determine the best OM. 
A range of simulations with $\chi^2$ within 1$\sigma$ of the minimum $\chi^2$ would be taken to determine the OM uncertainty from the analysis.
The $\chi^2$ value of each passing simulation was used in combination with the value of the parameters for that simulation to produce a weighted best fit for each parameter. An overall best fit, including covariance and correlations between parameters, was also produced from these passing simulations. 

\begin{figure}[h]
    \centering
    \includegraphics[width=0.68\textwidth]{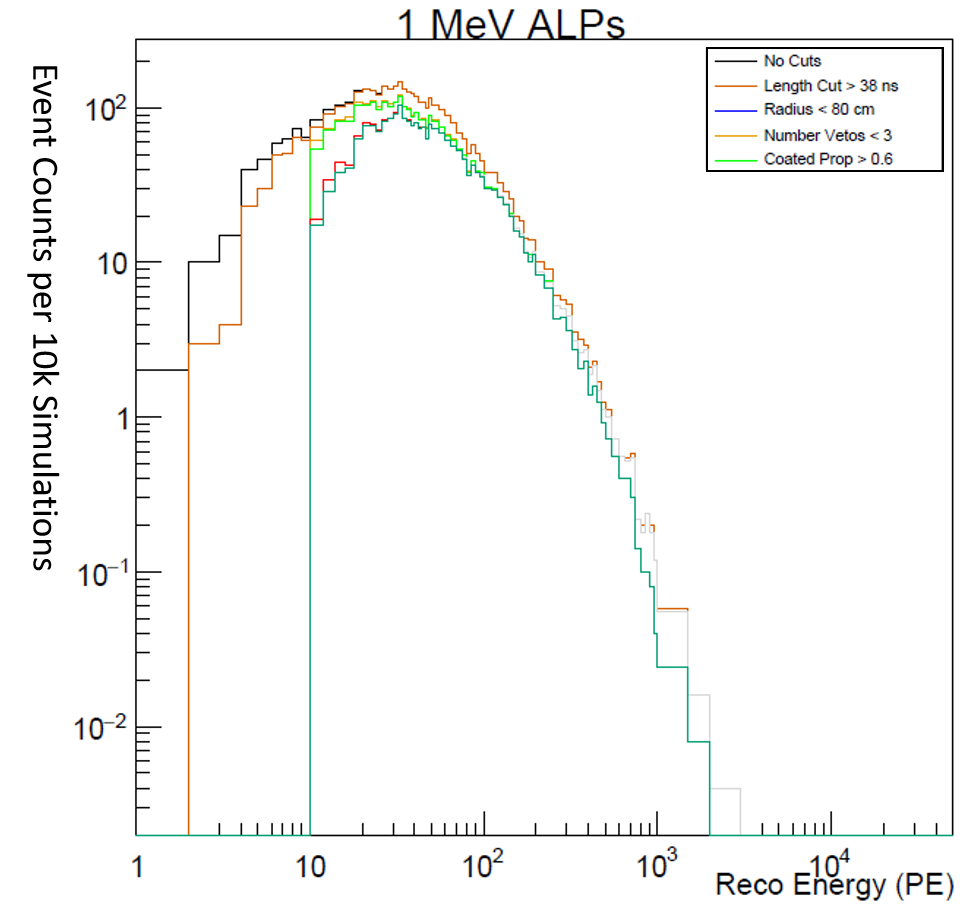}
    \caption{The reconstructed energy distribution from a large number of correlated OM simulations of a 1 MeV ALP event in CCM. The large width of the distribution represents the spread of outcomes from OM uncertainty, energy resolution, position resolution, and other sources of error. All these factors were encapsulated in the overall OM and OM covariance matrix, allowing the OM to be used for predicting signal efficiency and systematic uncertainty.}
    \label{fig:alp1mev}
\end{figure}

\begin{figure}[h]
    \centering
    \includegraphics[width=0.68\textwidth]{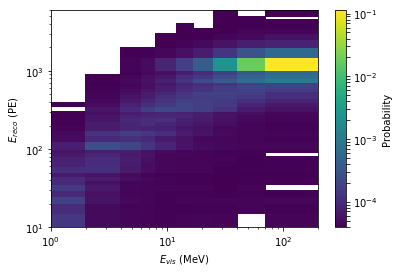}
    \caption{The generated smearing matrix for the ALP search across the expected potential ALP energies. The smearing matrix includes the efficiency for each energy, with the total for each column being equal to the total efficiency for that energy ALP.}
    \label{fig:alpSmearing}
\end{figure}

The generated optical model was used to determine the systematic error on potential signals.
To determine these systematic errors, we generated signal events - high energy gammas for ALP - across a range of energies and OM uncertainties. 
As shown in Figure \ref{fig:alp1mev}, the simulation OM produced a spread of results for the reconstructed energy even for a single true energy particle. 
This spread is a result of a number of factors, most notably significant differences in energy scale depending on the position inside the detector. 
However, this spread also includes the spread from OM uncertainty, as the OM parameters were varied in successive simulations according to the uncertainty from the best fit. 
We used the simulated events across the expected range of true energies - from 1 to 100 MeV - to build a smearing matrix (Figure \ref{fig:alpSmearing}) that would correlate between visible energy in the detector and reconstructed energy for the signal types. 
The smearing matrix was then applied to potential visible energy distributions as shown in Figure \ref{fig:spectra} to determine how a signal excess would be reconstructed. 
These reconstructed signals could be compared directly to the measured excess from the subtraction between ROI data and predicted background. 
Example excesses are shown overlaid onto the subtraction in Figure \ref{fig:subtract}.
Theses signal excesses include the OM uncertainty in the smearing matrix and thus the systematic error on potential signals.

These systematic errors ranged from 30\% at high energies to 50\% at low energies.
The primary contributors to this error are position variation, reconstruction resolution, and OM uncertainty. 
Position variation provides an uncertainty of 20-40\% depending on energy. 
Reconstruction resolution is $\pm10$ cm for position and $\pm15\%$ for energy, adding another 20\% uncertainty. 
Finally, OM uncertainty increases these uncertainties by around 10\%.
These uncertainties are added in quadrature to provide the final total systematic error on the signal.

\end{document}